\documentclass[11pt,psfig]{article}  
\usepackage{a4wide}
\usepackage{graphicx}
\usepackage{subfigure}
\usepackage[centertags]{amsmath}
\usepackage{amssymb}
\usepackage{dsfont}
\newcommand{\unity}{{ 1\:\!\!\!\mbox{I}}}

\newcommand{\Z}{{\mathbb Z}}
\newcommand{\R}{\mathcal R}
\newcommand{\OR}{\Omega {\mathcal R}}

\newcommand{\A}{\mbox{\bf \small A}}

\newcommand{\ku}{\mathcal{K}}
\newcommand{\au}{\mathcal{A}}
\newcommand{\msu}{\mathcal{M}}
\newcommand{\lu}{\mathcal{L}}
\newcommand{\kt}{\Tilde{\mathcal{K}}}
\newcommand{\at}{\Tilde{\mathcal{A}}}
\newcommand{\mt}{\Tilde{\mathcal{M}}}
\newcommand{\lt}{\Tilde{\mathcal{L}}}

\newcommand{\SC}{\scriptstyle}

\newcommand{\Sym}{\text{\bf S}}

\newcommand{\F}{\text{\bf F}}
\newcommand{\Fb}{\Bar{\text{\bf F}}}

\newcommand{\dd}{\text{d}}
\newcommand{\tr}{\mbox{tr}}

\newcommand{\barre}[1]{%
        \setbox1=\hbox{$#1$} \dimen2=\ht1 \dimen3=\dp1 \dimen4=\wd1  
        \setbox2=\hbox{\sl /}  
        \dimen1=\wd1 \advance\dimen1 by -\wd2 \divide\dimen1 by 2  
        \advance\dimen1 by \wd2 \advance\dimen1 by 0.4pt  
        \setbox3=\hbox to \wd1{\hss \box1 \kern -\dimen1 \box2\hss}  
        \ht3=\dimen2 \dp3=\dimen3 \wd3=\dimen4  
        \box3  
        }  
\begin{document}

\pagestyle{empty}  
\begin{flushright}  
                     hep-th/0201037 
\end{flushright}  
\vskip 2cm    

\begin{center}  
{\huge Intersecting brane world models from D8-branes on 
\mbox{$(T^2\times T^4/\Z_3 )/ \OR_1$} type IIA
  orientifolds}  
\vspace*{5mm} \vspace*{1cm}   
\end{center}  
\vspace*{5mm} \noindent  
\vskip 0.5cm

\centerline{{\bf Gabriele Honecker}}

\vskip 1cm
\centerline{\em Physikalisches Institut, Universit\"at Bonn}
\centerline{\em Nussallee 12, D-53115 Bonn, Germany}
\vskip2cm
  
\centerline{\bf Abstract}  
We present orientifold models of type IIA string theory with D8-branes
compactified
on a two torus times a four dimensional orbifold. The orientifold
group is chosen such that one coordinate of the two torus is reversed
when applying worldsheet parity. RR tadpole cancellation requires
\mbox{D8-branes}  which wrap 1-cycles on the two torus and transform
non-trivially under the orbifold group. These models are T-dual to
orientifolds with D4-branes only which admit large volume compactifications. The intersections of
the D8-branes are chosen in such a way that supersymmetry is broken in
the open string sector and chiral fermions arise. Stability of the
models is discussed in the context of NSNS tadpoles. Two examples with
the SM gauge group and two left-right symmetric models are given.

\vskip .3cm

  
\newpage  

\setcounter{page}{1} \pagestyle{plain}  

\section{Introduction} \label{intro}

If string theory is to be the underlying fundamental theory which
unifies gravity and the gauge interactions, low energy configurations
containing the standard model are expected to exist. For a long time,
$N=1$ supersymmetric compactifications on Calabi-Yau-threefolds  of
the ten dimensional heterotic string seemed to be the best
candidates. Some years ago, the situation started to change
due to the discovery of orientifold planes~\cite{Sagnotti:1987tw} and the
concept of D-branes~\cite{Polchinski:1995mt} as end points of open
strings which support the
gauge groups. With this discovery, the heterotic string lost its role
as the unique provider of phenomenologically interesting models. The
tools of obtaining the effective lower dimensional theories from
orientifold constructions were successively  worked
out~\cite{Pradisi:1989xd,Gimon:1996rq}. 

An appealing feature of type II orientifold constructions is the
possibility of explaining the hierarchy between the electroweak scale
$M_Z$ and the Planck scale $M_P$ in a natural way by taking the
internal volume transverse to the D-branes to be large.  By
this means, the string scale can be lowered down to the TeV
range~\cite{Arkani-Hamed:1998rs}.  

Four dimensional orientifold constructions of type II superstring
theories are determined by two major ingredients: Generically, the
compact space is considered to be an orbifold which is a  singular
limit of a Calabi-Yau-threefold or $K3$ times a two torus . Depending on the choice of the
orbifold, either a quarter or  half  of the original supersymmetries are
preserved. In addition, worldsheet parity in combination with some
spacetime action can be gauged. This gauging breaks another half of
the original supersymmetries. The geometric objects which are
invariant under the combined action of the orbifold and worldsheet
parity are called orientifold planes (O-planes). They can be
transversal to some of the compact dimensions and carry RR
charges. Consistency of the theory requires that RR charges are
cancelled by additional objects, the D-branes. They have the same
dimensionality as the O-planes. If the D-branes and O-planes are
situated on top of each other in the compact space, the
supersymmetry breaking is completely governed by the orbifold and
worldsheet parity. The resulting models have ${\cal N}=1$ or ${\cal
  N}=2$ supersymmetry in the closed string as well as the open string
sector. If on the other hand the D-brane data do not exactly match the
O-plane data, supersymmetry can be completely broken in the open
string sector while the closed sector remains ${\cal N}=1$ or ${\cal
  N}=2$ supersymmetric. Various realisations of the different scenarios have
been studied e.g. in~\cite{Dabholkar:1996zi,Gimon:1996ay,Berkooz:1997dw,Cvetic:2000hb,Blumenhagen:2000md}.

One possible way to realise the supersymmetry breaking in the open
string sector is given by including not only D-branes but also anti-D-branes
in the models. Strings stretching between a brane and an anti-brane
have masses depending on the distance between these branes. For small
distances, tachyons occur in the spectrum which render the theory
unstable and induce a phase transition~\cite{Banks:1995ch,Gaberdiel:2000jr}.

Another way of breaking supersymmetry in the open string sector of
type II orientifolds is to allow for a non-trivial magnetic background
flux in torus
compactifications~\cite{Bachas:1995ik,Angelantonj:2000hi}. These
fluxes can also trigger gauge and chiral symmetry breaking. In a
T-dual picture, the magnetic background fluxes are described by
relative angles of intersecting D-branes which wrap different cycles
in the compact space~\cite{Blumenhagen:2000wh,Hashimoto:1997gm}. In
this picture, 
massless chiral fermions are supported at the intersection loci of two
D-branes~\cite{Berkooz:1996km}. Tilting the compactification torus in
this picture corresponds to including a non-trivial discrete
NSNS-sector B field in the T-dual picture with magnetic background
fluxes~\cite{Angelantonj:2000rw,Blumenhagen:2001ea}. The effect of
this is a further gauge symmetry breaking as well as a change in the
number of generations.  
Such supersymmetry breaking four dimensional orientifold models with D6-branes at
angles on tori and some 
orbifolds have also been considered
in~\cite{Ibanez:2001nd,Rabadan:2001mt,Blumenhagen:2001te,Forste:2001gb}, and
in~\cite{Cvetic:2001tj} a particular subclass of such models was
discovered where by choosing special  intersection angles, chiral
symmetry is broken but supersymmetry is preserved.

In~\cite{Aldazabal:2001dg,Aldazabal:2001cn,Bailin:2001ie} the picture of branes at
angles was examined within the framework of type II superstring theory
on a four dimensional orbifold. Upon compactifying on an additional
two torus, D4-branes sitting at the orbifold fixed points can
intersect.

In this paper, we consider a hybrid ansatz of the previously discussed
ones. We construct orientifold models of type II superstring theory on
a four dimensional orbifold times a two torus. The orientifold group
is chosen such that D8-branes are required for tadpole
cancellation. These D8-branes wrap a non-trivial cycle on the two
torus while they fill the non-compact space and the orbifold. The
orbifold group acts non trivially on the Chan-Paton factors of the
open strings leading to a classification of representations of chiral
fermions according to the transformation properties of their
groundstates under the orbifold group. These models have a T-dual
description in terms of D4-branes and thus constitute the
orientifolded version of the models constructed
in~\cite{Aldazabal:2001dg} which admit large volume compactifications.  
One particular example of these models has already been presented
in~\cite{Honecker:2001dj}. While preparing the manuscript, the letter
article~\cite{Kataoka:2001sp} appeared where also intersecting
D4-branes in an unoriented theory are briefly mentioned.

The paper is organized as follows. In section~\ref{Construction}, we
present the construction of type II orientifold models with D8-branes
at angles. This includes the computation of RR tadpoles
and the generic chiral open spectrum. In section~\ref{sec_mixed_an}, we
comment on the Green-Schwarz mechanism required for cancellation of
mixed anomalies. In section~\ref{NSNStadpoles}, we discuss the
stability of our models in terms of NSNS tadpoles. In
section~\ref{examples}, we present two models with four generations and
the standard model gauge group as well as two left-right symmetric
models with three generations. The last
section~\ref{discussion}, is devoted to a discussion and
conclusions. Finally, details of the calculation are accumulated in the
appendices~\ref{app_1-loop}, \ref{app_tree}, \ref{app_massless_spectrum} and \ref{app_spectra_2a2b}.


\section{Construction of $(T^2\times T^4/Z_3)/\OR_1$ Orientifolds}
\label{Construction}

\subsection{General Setup}

In this article we discuss four dimensional orientifold models of type
IIA theory on $T^2\times T^4/Z_3$ with D8-branes at angles. The
four non-compact dimensions are labeled by $x^\mu$, $\mu = 0, \ldots
,3$. The compact space can be parameterized by three complex
coordinates,
\begin{equation}
  z^{1}=x^4 + ix^5 \, ,\, z^2 =x^6+ix^7 \, ,\, z^3 = x^8+ix^9 .
\end{equation}
corresponding to three two tori $T_{1,2,3}$. In the model under
consideration, worldsheet parity $\Omega$ is combined with a reflexion
on the first two torus,
\begin{equation}
  \R_1: z^1 \rightarrow \bar{z}^1 ,
\end{equation}   
and the generator $\Theta$ of $\Z_3$ acts on the second and third
torus, 
\begin{equation} 
  \Theta :\; z^i \rightarrow e^{2\pi iv_i} z^i,
\end{equation}
with $v=(0,1/3,-1/3)$. The sets of points which are left invariant
under $\OR_1\times \Theta$ constitute orientifold planes, which are
extended along all non-compact directions and the four dimensional
orbifold, but only along the $x^4$ axis on the first torus. Thus,
they extend along eight spatial dimensions. In order to
cancel the RR-charges of these O8-planes, an appropriate configuration
of D8-branes has to be added. Performing a T-duality along the $x^5$
direction, D8-branes at angles on $T_1$ correspond to D9-branes with
non-trivial magnetic background flux
$F$ which is quantized in terms of the radii of the two-torus, $F=\frac{\alpha'}{R_1R_2}\frac{q}{p}$~\cite{Hashimoto:1997gm,Blumenhagen:2000wh}. In addition, toroidal
compactifications of type I string theory allow for a non-trivial
constant background NSNS two-form flux
$b=0,1/2$~\cite{Bianchi:1992eu}. In the T-dual picture with D8-branes at angles,
the discrete value $b=1/2$ corresponds to a tilted torus lattice
w.r.t. the real axis as discussed in~\cite{Blumenhagen:2001ea} or the
{\bf b} type lattice of~\cite{Forste:2001gb}. The relation between
these two ways of describing the torus is explicitly shown in
appendix~\ref{app_lattice_con_T2}. D$8_a$-branes are specified by the
wrapping numbers $(n_a,m_a)$ along the two fundamental cycles of
$T_1$. In the T-dual picture with D9-branes,
these wrapping numbers are replaced by the quantization condition on
the magnetic and the NSNS background flux, i.e. $(p_a,q_a)\simeq (n_a,m_a+bn_a)$. 
Due to the reflexion symmetry ${\cal R}_1$, each D$8_a$-brane is
accompanied by its mirror image D$8'_a$ with wrapping numbers 
$(n'_a,m'_a)=(n_a,-m_a-2bn_a)$. Two stacks of branes D$8_a$ and
D$8_b$ generically have several intersections within the fundamental
cell of the torus. The
corresponding intersection numbers can be expressed in terms of the
wrapping numbers,
\begin{equation}\label{intersect_number}
I_{ab} = n_a m_b - n_b m_a.
\end{equation}
Formally, the intersection number can take negative values. In terms
of physical quantities, this means that the particles with support at
the intersection locus transform under the conjugate representation.
In the model under consideration, the orbifold generator $\Theta$
preserves the position of each D$8_a$ brane while assigning different
phases $\alpha^j$ (where $\alpha \equiv e^{2\pi i/3}$ and
$j=0,1,2$). Therefore, a stack of $N_a$ branes with identical positions
is decomposed according to the different eigenvalues of the $\Z_3$
rotation, $N_a=N^0_a+N^1_a+N^2_a$,  giving rise to the gauge
group
\begin{equation}
U\left(N^0_a \right) \times U\left(N^1_a \right)\times U\left(N^2_a \right).
\end{equation}
Particles which are supported at the intersection locus of two stacks
of branes D$8_a$ and D$8_b$ with $\Z_3$ eigenvalue 1 transform as
$\left(N^i_a, \Bar{N}^i_b \right)$ whereas those with eigenvalue
$\alpha^{\pm 1}$ transform as  $\left(N^i_a, \Bar{N}^{i\pm 1}_b \right)$. 

The gauge coupling constants of the $U(N^i_a)$ factors with support on
a D$8_a$ brane are determined by the length $L_a$ of the 1-cycle on
$T_1$ which the D$8_a$ brane wraps~\cite{Aldazabal:2001cn},
\begin{equation}
\frac{2\pi}{g_a^2}\sim\frac{M_s}{\lambda_s}L_a. 
\end{equation}
The length of the cycle in terms of wrapping numbers and radii of
the two-torus is given by
\begin{equation}
L_a=\sqrt{(n_aR_1)^2+((m_a+bn_a)R_2)^2}.
\end{equation}

In this class of models, Yukawa couplings arise from bosonic and
fermionic fields living at the intersection loci of three different types of
D$8_{a,b,c}$ branes. These intersection points constitute the cusps of a
triangle of area $A_{abc}$ (which is dimensionless when the area of
the torus $T_1$ is measured in string units). The size of Yukawa couplings is
exponentially suppressed in terms of the area~\cite{Aldazabal:2001cn},
\begin{equation}\label{yukawa}
Y_{abc}=\exp\left(-A_{abc}\right).
\end{equation}

\subsection{RR-Tadpole Cancellation}
\label{section_RR_tadpoles}

In this section, we derive the consistency conditions of the ($T^2
\times T^4/\Z_3)/\OR_1$ models which are determined by the requirement that
all -- untwisted and twisted -- RR-charges of the O8-planes are
cancelled by those of the D$8_a$-branes. These so called tadpole
cancellation conditions can be entirely expressed in terms of the
wrapping numbers $n_a$ corresponding to the projection onto the $x^4$-axis and the number of identical
branes $N^i_a$. 

\begin{figure}
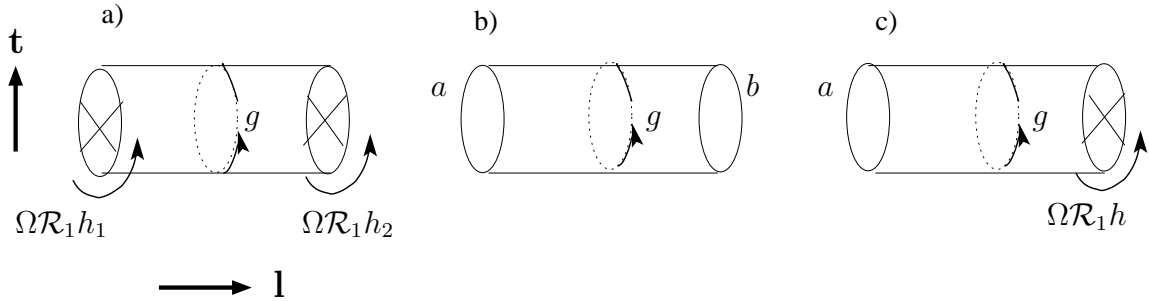
  
\begin{center}
\input amplitudes.pstex_t
\end{center}
\caption{a) Klein bottle, b) Annulus, c) M\"obius strip}
\label{fig_amplitudes}
\end{figure}

In the closed string sector at 1-loop, the Klein-bottle amplitude
contributes to the RR-tadpole. This divergence can be cancelled by the
two 1-loop amplitudes of the open string sector, namely the annulus
and the M\"obius strip.  The 1-loop amplitudes are depicted
in figure~\ref{fig_amplitudes} where time $t$ evolves along the vertical
direction. By choosing the time $l$ to run along the horizontal
direction instead, one obtains the picture in the tree-channel where
the Klein bottle amplitude is given by a closed string scattered
between two O8-planes, the annulus by scattering between a D$8_a$ and
a D$8_b$-brane and the M\"obius strip between a D$8_a$-brane and an O8-plane.
The O8-planes are described by crosscap states invariant under
$\OR_1h$ where $g$, $h$ label elements of the orbifold group
$\Z_N$. $g$ denotes the twist sector of the closed string propagating
in the tree-channel. Consistency of the boundary conditions
requires in the Klein bottle diagram~\cite{Gimon:1996rq} 
\begin{equation}
\left(\OR_1h_1\right)^2=\left(\OR_1h_2\right)^2=g
\end{equation}
and in the M\"obius strip
\begin{equation}
\left(\OR_1h\right)^2=g.
\end{equation}
In contrast to the models with branes at angles considered
in~\cite{Blumenhagen:2000md,Blumenhagen:2001te,Forste:2001gb},
the following relation holds
\begin{equation}
\left(\OR_1h\right)^2=h^2.
\end{equation}
Therefore, twisted as well as untwisted closed strings
propagate in the tree channel leading to untwisted and twisted tadpole
cancellation conditions which have to be fulfilled simultaneously. 

At this point, we turn to the explicit calculation of the three
1-loop-amplitudes. The direct calculation in the tree-channel can be
performed using the boundary state approach (see e.g.~\cite{Gaberdiel:2000jr}). For our class of
models, the relevant formulas are displayed in appendix~\ref{app_tree}.
The constraints on $N^i_a$ can, however, only be read off by starting
from the 1-loop amplitudes.

\subsubsection{Klein bottle}

The closed string 1-loop contribution to the RR-exchange in the
tree-channel can be obtained by computing the
NSNS part with $(-1)^F$ insertion (where $F$ is the worldsheet fermion
number). The lattice contributions ${\cal L}_1$ on $T_1$ where the reflexion 
${\cal R}_1$ acts are as discussed
in~\cite{Blumenhagen:2001ea,Forste:2001gb}. In addition, in the
untwisted sector Kaluza Klein momenta arise along all directions of
the orbifold whereas windings are projected out by worldsheet
parity. The explicit formulas for the lattice contributions of the
orbifold  to the
amplitudes are listed in appendix~\ref{app_lattice_con_T4}. $\OR_1$
exchanges $\Theta$ and $\Theta^{-1}$ twisted sectors. Hence, in the
1-loop channel, only untwisted sectors contribute. The calculation of
the contribution with $\unity$ insertion goes completely along the
lines discussed in~\cite{Blumenhagen:2001ea,Forste:2001gb} yielding  
\begin{equation}\label{kuU}
\ku_U =\frac{c}{3}\int_0^{\infty}\frac{\dd t}{t^3} \lu^{\ku}_{1}
\lu^{\ku}_{2}\lu^{\ku}_{3}\ku^{(0)}
\end{equation}
where $c \equiv V_4/(8\pi^2\alpha^{\prime})^2$ contains the
regularized four dimensional volume $V_4$ arising from
integrating out the non-compact momenta. $\lu^{\ku}_{2}\lu^{\ku}_{3}$
is as given in~(\ref{app_latt_kb}) in appendix~\ref{app_lattice_con_T4}. Performing the modular
transformation $t=1/(4l)$ gives the contribution from the untwisted
RR-fields,
\begin{equation}\label{ktU}
\ku^U =\frac{c}{3}\int_0^{\infty} \dd l \frac{256}{3} \frac{R_1}{R_2}
 \omega  \lt^{\ku}_1\lt^{\ku}_2\lt^{\ku}_3 \kt^{(0)}.
\end{equation}
where $R_{1,2}$ are the two radii of the first two-torus $T_1$ and $\omega$
is the volume of the orbifold $T^4/\Z_3$.  

In addition, $\Theta^{1,2}$ insertions create tadpoles which are
independent of the internal volume of the orbifold,
\begin{equation}\label{kuT}
\ku^T =\frac{c}{3}\int_0^{\infty}\frac{\dd t}{t^3} \lu^{\ku}_{1}
\sum_{k=1}^{2}\ku^{(k)}.
\end{equation}
The explicit expression of $\ku^{(k)}$ in terms of generalized
Jacobi-Theta functions is given in formula (\ref{app_osc_kb_u}) in
appendix~\ref{app_osc}. The
lattice contributions $\lu^{\ku}_{1}$ are the same as in formula
(\ref{kuU}), whereas the Kaluza Klein momenta on $T_{2,3}$ are not
invariant under $\Theta$. Transforming to the tree channel, the
twisted Klein bottle is given by
\begin{equation}\label{ktT}
\ku^T =-16\frac{c}{3}\int_0^{\infty} \dd l
\frac{R_1}{R_2}\lt^{\ku}_1\sum_{k=1}^2 \kt^{(k)}
\end{equation}
where $\kt^{(k)}$ is given by (\ref{app_osc_kb_t}) in appendix~\ref{app_osc}.

\subsubsection{Annulus}

The annulus amplitude is obtained from open strings stretching between
branes D$8_a$  and D$8_b$ at angle $\pi \Delta \varphi_{ab}$ on $T_1$. The
contributions from $T_1$ have been discussed in detail 
in~\cite{Blumenhagen:2001ea,Forste:2001gb}. The computation of the
trace with trivial insertion is again completely analogous to the one
performed in~\cite{Blumenhagen:2001ea,Forste:2001gb} yielding the
untwisted RR-tadpole of the annulus in the tree-channel
\begin{equation}\label{atU}
 \au^U_{ab} = -N_aN_bI_{ab}\frac{c}{3}\int_0^{\infty} \dd l 
 \frac{1}{6}\omega \at^{(0)}\lt^{\au}_{2}\lt^{\au}_{3}.
\end{equation}
$N_a$ labels the number of D$8_a$ branes of identical position,
$I_{ab}$ is the intersection number on $T_1$ defined in
(\ref{intersect_number}), $\lt^{\au}_{2}\lt^{\au}_{3}$ is given in
(\ref{app_latt_an}) in appendix~\ref{app_lattice_con_T4} and the oscillator contribution is given by
\begin{equation}
\at^{(0)}=\frac{\vartheta \Bigl[\!\! \begin{array}{c} \SC 1/2 \\ \SC 0
    \end{array} \!\!\Bigr]^3}{\eta^{9}}
 \frac{\vartheta \Bigl[\!\! \begin{array}{c} \SC 1/2 \\ \SC \Delta\varphi
    \end{array} \!\!\Bigr]}
  {\vartheta \Bigl[\!\! \begin{array}{c} \SC 1/2 \\ \SC 1/2+\Delta\varphi
    \end{array} \!\!\Bigr]}(2l)
\stackrel{l\rightarrow \infty}{\longrightarrow}
\frac{-8}{ I_{ab}}
\left(n_an_b\frac{R_1}{R_2}+(m_a+bn_a)(m_b+bn_b)\frac{R_2}{R_1}\right).
\end{equation}
The explicit dependence of the annulus tadpole on the 
orbifold volume $\omega$ is due to the fact that D$8_a$ branes have
Neumann directions along $x^{6\ldots9}$ leading to Kaluza Klein
momenta $p^{6\ldots9}$.

In addition to the trivial insertion, each $\Theta^k$ insertion
preserves the positions of branes. Kaluza Klein momenta are projected
out, and the $\Z_3$ rotation acts non-trivially on the Chan-Paton
labels of the open string with endpoints on branes $a,b$ via the
matrices $\gamma^a_{\Theta^k}, \gamma^b_{\Theta^k}$ leading to 
\begin{equation}\label{auT}
 \au^T_{ab} =  \frac{I_{ab}}{4}\frac{c}{3}\int_0^{\infty}\frac{\dd t}{t^3}
 \sum_{k=1}^2 \tr\gamma^a_{k}\tr\gamma^{-1,b}_{k}\au^{(k)}
\end{equation}
with $\au^{(k)}$ explicitly listed in (\ref{app_osc_an_u}) in appendix~\ref{app_osc}. By modular
transformation $t=1/(2l)$, one arrives at the twisted RR-tadpole
contribution of the annulus,
\begin{equation}\label{atT}
 \au^T_{ab} =  -I_{ab}\frac{c}{3}\int_0^{\infty} \dd l \frac{1}{2}
 \sum_{k=1}^2 \tr\gamma^a_{k}\tr\gamma^{-1,b}_{k}\at^{(k)}
\end{equation}
with $\at^{(k)}$ given by (\ref{app_osc_an_t}) in appendix~\ref{app_osc}. Thus, the asymptotic
behaviour of the annulus amplitudes is given by
\begin{align}
 \au^U_{ab} &\stackrel{l\rightarrow \infty}{\longrightarrow} 
N_aN_b\frac{4}{3}\omega\frac{c}{3}\int_0^{\infty}\dd l
\left(n_an_b\frac{R_1}{R_2}+(m_a+bn_a)(m_b+bn_b)\frac{R_2}{R_1}\right),\\
 \au^T_{ab} &\stackrel{l\rightarrow \infty}{\longrightarrow} 
- \frac{c}{3}\int_0^{\infty}\dd l
\sum_{k=1}^2 \tr\gamma^a_{k}\tr\gamma^{-1,b}_{k}
\left(n_an_b\frac{R_1}{R_2}+(m_a+bn_a)(m_b+bn_b)\frac{R_2}{R_1}\right).
\end{align}

\subsubsection{M\"obius strip}

The computation of the untwisted RR-exchange in the tree channel arising from
the M\"obius strip amplitude is again very similar to the case
discussed in~\cite{Blumenhagen:2001ea,Forste:2001gb}. Only strings stretching between
mirror branes $a$ and $a'$ contribute. Their multiplicity is
determined by the number of $\OR_1$ invariant intersections
$I^{\OR_1}_{aa^{\prime}}$ listed in
appendix~\ref{app_intersection_numbers}, and the Neumann directions
on $T_{2,3}$ lead to lattice contributions from Kaluza Klein momenta
displayed in (\ref{app_latt_ms}) in appendix~\ref{app_1-loop}. Therefore, the untwisted RR-exchange
is linearly proportional to the orbifold volume $\omega$. 

The computation of the twisted RR-tadpoles in the M\"obius strip is
also completely 
analogous to the annulus case. The $\Z_3$ rotation acts non-trivially on
the Chan-Paton-matrix of the $aa'$ string, lattice contributions are
projected out and the oscillator contributions are listed in
(\ref{app_osc_ms_t}) in appendix~\ref{app_osc}.  In summary, we obtain the asymptotic behaviour
\begin{align}\label{mstU}
\msu^U_a &\stackrel{l\rightarrow \infty}{\longrightarrow} 
-\frac{c}{3}\int_0^{\infty}\dd l
\frac{256}{3}
\frac{R_1}{R_2}\omega
n_a \tr\left(\gamma^{-1,a'}_{\OR_1}\gamma^{T,a}_{\OR_1}\right),\\
\label{mstT}
\msu^T_a &\stackrel{l\rightarrow \infty}{\longrightarrow} 
\frac{c}{3}\int_0^{\infty}\dd l
16n_a\frac{R_1}{R_2}\sum_{k=1}^{2} 
\tr\left(\gamma^{-1,a'}_{\OR_1k}\gamma^{T,a}_{\OR_1k}\right).
\end{align}
The trace in (\ref{mstT}) can be transformed due to closure of the
orientifold group, i.e.
\begin{equation}
\gamma_{k+l}^a =c_{k+l}
\gamma^{\prime-T,a'}_{\OR_1l}\gamma^a_{\OR_1k}.  
\end{equation}

\subsubsection{RR-tadpole cancellation}

The RR-tadpole cancellation conditions can be extracted from the
asymptotic behaviour of the Klein bottle ((\ref{ktU}) and (\ref{ktT})), the
annulus ((\ref{atU}) and (\ref{atT})) and the M\"obius strip
((\ref{mstU}) and (\ref{mstT})) after summing over all possible open
string configurations. 

The untwisted tadpole conditions are
\begin{eqnarray}
  \left[\sum_a n_a N_a-16\right]^2=0,\label{tad_con_U_1}\\
  \tr\left(\gamma^{-1,a'}_{\OR_1}\gamma^{T,a}_{\OR_1}\right)=N_a.\label{tad_con_U_2} 
\end{eqnarray}
The twisted tadpole conditions split into the projection onto the
$x^4$ axis proportional to $R_1/R_2$ and to the $x^5$ direction
proportional to $R_2/R_1$,
\begin{eqnarray}
\frac{R_2}{R_1}: \qquad \label{tad_con_T_1}&
\sum_{k=1}^{2}\Big|\sum_{a}(m_a+bn_a)
\left(\tr\gamma^a_{k}-\tr\gamma^{a'}_{k}\right)\Big|^2=0,\\
\frac{R_1}{R_2}: \qquad \label{tad_con_T_2}&
\sum_{k=1}^{2}\Bigl(
8^2+\Big|\sum_{a}n_a\left(\tr\gamma^a_{k}+\tr\gamma^{a'}_{k}\right)\Big|^2
-2\cdot 8 \cdot 
\sum_{a}n_a\left(c_{2k}\tr\gamma^a_{2k}
+\Tilde{c}_{2k}\tr\gamma^{a'}_{2k}\right)
\Bigr)=0.
\end{eqnarray}
Condition (\ref{tad_con_T_1}) is trivially fulfilled if for mirror
branes D$8_a$ and D$8_{a'}$ the identity $\tr\gamma^a_{k}= \tr\gamma^{a'}_{k}$
holds. Furthermore, equation (\ref{tad_con_T_2})  gives a total square
for each twist sector $k$ provided that $c_{2k}=\Tilde{c}_{2k}=1$ and
$\tr\gamma_{2k}\in {\mathbb R}$.  These conditions fix the form of
$\gamma^a_{\Theta}$,
\begin{equation}\label{rep_gamma_k}
\gamma^a_{\Theta}=\text{diag}\left(\unity_{N_a^0}, e^{2\pi i/3}_{N_a^1}, 
e^{-2\pi i/3}_{N_a^2}\right) 
\end{equation}
with $N_a=N^0_a+N^1_a+N^2_a$ and $N^1_a=N^2_a$. 

Inserting (\ref{rep_gamma_k}) in (\ref{tad_con_U_1}) and
(\ref{tad_con_T_2}) determines the RR-tadpole cancellation conditions
entirely in terms of the wrapping numbers $n_a$ along the $x^4$ axis
and the number of identical branes $N^i_a$,
\begin{eqnarray}
\sum_a n_a N^0_a=8,\label{tad_con_A1}\\
\sum_a n_a N^1_a=4.\label{tad_con_A2}
\end{eqnarray}
So far, we have only considered D$8_a$ branes which are mapped to
their mirror image D$8_{a'}$ under the reflexion ${\cal R}_1$. A D$8_c$
brane which is its own mirror image contributes only half the amount
to the tadpole conditions, i.e.
\begin{eqnarray}
\frac{n_cN^0_c}{2}+\sum_{a\neq c} n_a N^0_a &= 8,\label{tad_con_N1}\\
\frac{n_cN^1_c}{2}+\sum_{a\neq c} n_a N^1_a &= 4.\label{tad_con_N2}
\end{eqnarray}
The wrapping numbers of the $\OR_1$ invariant brane are
$(n_c,m_c)=(1,0)$ for vanishing background antisymmetric NSNS tensor field
$b$ and $(n_c,m_c)=(2,-1)$ for $b=1/2$. In the limit
\mbox{$R_1,\frac{1}{R_2}\rightarrow \infty$} where the T-dual two-torus $T_1$ decompactifies,
the supersymmetric six dimensional set-up is recovered which for
vanishing antisymmetric NSNS tensor, 
i.e. a single stack of branes with $(n_c,m_c)=(1,0)$ and $b=0$, is
identical to the $\Z_3$ orientifold in~\cite{Gimon:1996ay}.

\subsection{Chiral open spectrum}
\label{section_chiral_spectrum}

The closed string spectrum contains the ${\cal N}=2$ SUGRA multiplet
as well as eleven hypermultiplets and ten tensor multiplets. The
complete closed string sector is ${\cal N}=2$ supersymmetric and
non-chiral. 
In order to determine the open string spectrum, we fix the Chan-Paton
matrices 
\begin{align}
\gamma^a_{\OR_1} = \gamma^{a'}_{\OR_1} &= \left(
\begin{array}{ccc} \unity_{N_a^0} & 0 & 0 \\
    0 & 0 & \unity_{N_a^1}\\
    0 & \unity_{N_a^1} & 0 \end{array} \right),\\
\gamma^a_{\Theta} = \gamma^{a'}_{\Theta} &= \text{diag}\left(\unity_{N_a^0}, e^{2\pi i/3}_{N_a^1}, 
e^{-2\pi i/3}_{N_a^1}\right)
\end{align}
in analogy to the supersymmetric case discussed
in~\cite{Gimon:1996ay}. Open strings stretching between identical
branes then give the gauge groups
\begin{equation}
U(N_a^0)\times \left(U(N_a^1)\right)^2.
\end{equation}
In the case of an $\OR_1$ invariant brane, the gauge group is reduced
to 
\begin{equation}
SO(N_a^0)\times U(N_a^1).
\end{equation}
The $aa$ sectors of open strings are again ${\cal N}=2$ supersymmetric and
non-chiral. 

Finally, the sector of $ab$ strings stretching between D$8_a$ and D$8_b$ branes
at angles $\pi \Delta \varphi_{ab}$ is non-supersymmetric and
chiral. This part of the spectrum generically contains
tachyons since masses of states in the NS sector are given by
\begin{equation}
\alpha^{\prime}m_{ab}^2=\text{osc}+\frac{\Delta \varphi_{ab}}{2}-\frac{1}{2}.
\end{equation}
Thus, the state $\psi^{1}_{\Delta\varphi-1/2}|0\rangle_{NSNS}$ is
tachyonic. A complete list of lightest NS states is given in
appendix~\ref{app_massless_spectrum}. In contrast to the models with
D$6_a$ branes discussed in~\cite{Forste:2001gb}, mass eigenstates in
the models with D$8_a$ branes have to be classified according to their
$\Z_3$ eigenvalue. Tachyonic states only occur in the sectors with
eigenvalue 1. In principle, this introduces the possibility of
choosing the brane set-up, i.e. the numbers $N^i_a$, such that no
chiral sector with trivial eigenvalue occurs. However, the tadpole
conditions~(\ref{tad_con_N1}), (\ref{tad_con_N2}) constrain the models
severely. Furthermore, in contrast to the type IIB models examined
in~\cite{Aldazabal:2001dg,Aldazabal:2001cn,Bailin:2001ie} the
orientifold projection $\OR_1$ enforces the existence of mirror
branes. $aa^{\prime}$ strings automatically include a sector
containing tachyons which can be only projected out completely by the $\OR_1$
symmetry in case of a single $U(1)_a$ gauge factor.    

The R sector of branes at angles provides chiral fermions. The
groundstate ist fourfold degenerated as displayed in the table in
appendix~\ref{app_massless_spectrum}. The degeneracy is lifted by the
$\Z_3$ symmetry. In summary, the chiral spectrum is listed in
table~\ref{tab_gen_spect}. For an $\OR_1$ invariant brane D$8_c$, the
spectrum is slightly changed as displayed in
table~\ref{tab_gen_spect_OR_inv}. Generically, the sector $a'b'$
provides the anti-particles of the $ab$ sector and the sector $ab'$ is
paired with $a'b$. But for $c=c'$, only the sectors $cb$ and $cb'$ are
present and form a pair. RR-tadpole cancellation ensures that the chiral spectrum is
free of purely non-abelian gauge anomalies. Mixed $U(1)$ anomalies
will have to be cured by a generalized Green-Schwarz mechanism
involving twisted RR-fields from the closed string
sector~\cite{Douglas:1996sw,Aldazabal:2001dg,Lalak:1999bk}. 

The chiral $aa'$, $ab$ and $ab'$ sectors with $\Z_3$ eigenvalue 1 are accompanied
by a tachyonic scalar. As already mentioned in the previous paragraph,
the $aa'$ sector is only absent provided that $n_a=1$ and $N^0_a=1,
N^1_a=N^2_a=0$, i.e. the D$8_a$ brane accommodates a single $U(1)_a$
gauge factor. 

\renewcommand{\arraystretch}{1.3}
\begin{table}[ht]
  \begin{center}
    \begin{equation*}
      \begin{array}{|cl||c|c|c|} \hline
        \multicolumn{4}{|c|}{\rule[-3mm]{0mm}{8mm} \text{\bf
Massless chiral fermionic spectrum on $T^2 \times T^4/\Z_3$ with D8-branes}}
\\\hline\hline 
\text{sector} & \Z_3  & \text{multiplicity} & \text{rep.}
\\\hline\hline 
        aa'  & 1 & 2(2m_a+(2b)n_a) 
        & (\A_a^0,1,1) + (1,\F_a^1,\F_a^2)\\
        &&(n_a-1)(2m_a+(2b)n_a) 
        & (\A_a^0+\Sym_a^0,1,1) + 2(1,\F_a^1,\F_a^2)\\
        & \alpha & (2m_a+(2b)n_a) 
        & (\Fb_a^0,1,\Fb_a^2)+(1,\Bar{\A}_a^1,1)\\
        && \frac{n_a-1}{2}(2m_a+(2b)n_a) 
        & 2(\Fb_a^0,1,\Fb_a^2)+(1,\Bar{\A}_a^1+\Bar{\Sym}_a^1,1)\\
        & \alpha^2 & (2m_a+(2b)n_a) 
        & (\Fb_a^0,\Fb_a^1,1)+(1,1,\Bar{\A}_a^2)\\
        && \frac{n_a-1}{2}(2m_a+(2b)n_a) 
        & 2(\Fb_a^0,\Fb_a^1,1)+(1,1,\Bar{\A}_a^2+\Bar{\Sym}_a^2)\\\hline
        ab & 1 & 2(n_am_b-n_bm_a)
        &(\Fb_a^0,\F_b^0)+(\Fb_a^1,\F_b^1)+(\Fb_a^2,\F_b^2)\\
        & \alpha &(n_am_b-n_bm_a) 
        &(\F_a^0,\Fb_b^1)+(\F_a^1,\Fb_b^2)+(\F_a^2,\Fb_b^0)\\
        & \alpha^2 & (n_am_b-n_bm_a)
        &(\F_a^0,\Fb_b^2)+(\F_a^1,\Fb_b^0)+(\F_a^2,\Fb_b^1)\\\hline
        ab' & 1 & 2(n_am_b+n_bm_a+(2b)n_an_b)
        &(\F_a^0,\F_b^0)+(\F_a^1,\F_b^2)+(\F_a^2,\F_b^1)\\
        & \alpha &(n_am_b+n_bm_a+(2b)n_an_b) 
        &(\Fb_a^0,\Fb_b^2)+(\Fb_a^1,\Fb_b^1)+(\Fb_a^2,\Fb_b^0)\\
        & \alpha^2 & (n_am_b+n_bm_a+(2b)n_an_b)
        &(\Fb_a^0,\Fb_b^1)+(\Fb_a^1,\Fb_b^0)+(\Fb_a^2,\Fb_b^2)\\\hline
      \end{array}
    \end{equation*}
  \end{center}
\caption{Chiral spectrum. The sectors are classified by the $\Z_3$
        eigenvalue of the corresponding R groundstate.}
\label{tab_gen_spect}
\end{table}

\renewcommand{\arraystretch}{1.3}
\begin{table}[ht]
  \begin{center}
    \begin{equation*}
      \begin{array}{|cl||c|c|c|} \hline
        \multicolumn{4}{|c|}{\rule[-3mm]{0mm}{8mm} \text{\bf
Chiral fermions for an $\OR_1$ invariant brane D$8_c$}}
\\\hline\hline 
\text{sector} & \Z_3  & \text{multiplicity} & \text{rep.}
\\\hline\hline 
        cb & 1 & 2n_c(m_b+bn_b)
        &(\Fb_c^0,\F_b^0)+(\Fb_c^1,\F_b^1)+(\F_c^1,\F_b^2)\\
        & \alpha &n_c(m_b+bn_b)
        &(\F_c^0,\Fb_b^1)+(\F_c^1,\Fb_b^2)+(\Fb_c^1,\Fb_b^0)\\
        & \alpha^2 & n_c(m_b+bn_b)
        &(\F_c^0,\Fb_b^2)+(\F_c^1,\Fb_b^0)+(\Fb_c^1,\Fb_b^1)\\\hline
      \end{array}
    \end{equation*}
  \end{center}
\caption{Modification of the chiral spectrum involving an $\OR_1$ invariant brane $c$.}
\label{tab_gen_spect_OR_inv}
\end{table}

\section{Cancellation of Mixed Anomalies}
\label{sec_mixed_an}

The generic chiral open spectrum displayed in
table~\ref{tab_gen_spect} and~\ref{tab_gen_spect_OR_inv} is free of purely non-abelian gauge
anomalies, but yields mixed gravitational anomalies of the form 
\begin{equation}\label{mix_anom_1}
U(1)_{i,a} - g_{\mu\nu}: 
6\left(2\delta_{i,0}-\delta_{i,1}-\delta_{i,2}\right) \left(m_a+bn_a\right)N^i_a
\end{equation}
as well as mixed gauge anomalies which for $(i,a)\neq (j,b)$ are proportional to  
\begin{eqnarray}\label{mix_anom_2}
U(1)_{i,a} - SU(N^j_b)^2:
\Bigl\{ & \left(m_a+bn_a\right)n_b
  \left(2\delta_{i,0}-\delta_{i,1}-\delta_{i,2}\right) 
\left(2\delta_{j,0}-\delta_{j,1}-\delta_{j,2}\right)\Bigr.\\
\Bigl. & -3 n_a \left(m_b+bn_b\right)\left(\delta_{i,1}-\delta_{i,2}\right)
\left(\delta_{j,1}-\delta_{j,2}\right) 
\Bigr\}N^i_a C_2(N^j_b)\nonumber
\end{eqnarray}
where $C_2(N)=\frac{N^2-1}{2N}$
is the quadratic Casimir of the fundamental representation of $SU(N)$.

Consistency of the models requires anomalous gauge fields to acquire a
mass and thus decouple from the effective low energy field
theory. This is realized by the Green-Schwarz mechanism which
in models with $K3$ orbifold compactifications involve twisted sector
fields~\cite{Douglas:1996sw}. The potential
candidates are the RR scalars ${}^{6}C^{(0)}_k$ and two-forms
${}^{6}C^{(2)}_k$ in six dimensions which belong to the sixdimensional
twisted hyper- and tensormultiplets, respectively. They arise from
the Kaluza Klein reduction of the ten-dimensional two form
${}^{10}C^{(2)}$ and self-dual four form ${}^{10}C^{(4)}$ on a vanishing
supersymmetric two-cycle $\Sigma_k$ on the orbifold,
\begin{equation}
{}^{6}C^{(2)}_k=\int_{\Sigma_k}{}^{10}C^{(4)},
\qquad
{}^{6}C^{(0)}_k=\int_{\Sigma_k}{}^{10}C^{(2)}.
\end{equation}
The scalar has a dual four form in six dimensions,
\begin{equation}
{}^{6}C^{(4)}_k=\int_{\Sigma_k}{}^{10}C^{(6)}.
\end{equation}
Modding out the worldsheet parity amounts to mapping different cycles
$\Sigma_k$ onto each other such that for the $T^4/\Z_3$ limit, $k$
runs over nine distinct values.

Reducing further down to four dimensions, the pullback of a closed
sector two form
on a multiply wrapped brane gives a scalar times the wrapping number
along the $\OR_1$ invariant direction~\cite{Aldazabal:2001dg},
\begin{equation}
n_bB^{(0)}_k=\int_{T^2(D9_b)}{}^{6}C^{(2)}_k
\qquad
n_bB^{(2)}_k=\int_{T^2(D9_b)}{}^{6}C^{(4)}_k
\end{equation}
while integrating out the two form ${\cal F}_a=F_a+B_a
=\frac{(m_a+bn_a)\alpha'}{n_aR_1R_2}$ on the torus yields
  as prefactor $m_a+bn_a$. 
The resulting four dimensional couplings are of the form
\begin{eqnarray}\label{GS_mech}
\left(m_a+bn_a\right) \int_{R^{1,3}} \text{tr}\left(\gamma^a_k\lambda^a_i\right)
C^{(2)}_k \wedge  F_{a,i}
\qquad&  &\qquad
n_b\int_{R^{1,3}}\text{tr}
\left(\gamma^b_k\lambda^b_i\lambda^b_j\right) B^{(0)}_k F_{b,i} \wedge
F_{b,j}\\
n_a \int_{R^{1,3}} \text{tr}\left(\gamma^a_k\lambda^a_i\right)
B^{(2)}_k \wedge  F_{a,i}
\qquad& &\qquad
\left(m_a+bn_a\right) \int_{R^{1,3}}\text{tr}
\left(\gamma^b_k\lambda^b_i\lambda^b_j\right) C^{(0)}_k F_{b,i} \wedge
F_{b,j}\nonumber
\end{eqnarray}
where $\lambda^a_i$ is the Chan-Paton-factor belonging to the
gauge-field component $F_{a,i}$. 

\begin{figure}
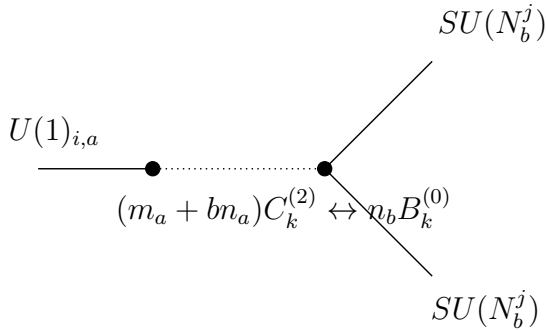
  
\begin{center}
\input mixed_gauge_an.pstex_t
\end{center}
\label{pic_GS_mech}
\caption{Green-Schwarz counter terms}
\end{figure}
The expressions on the left hand side in~(\ref{GS_mech}) render the anomalous gauge fields
massive. Combining the two couplings ~(\ref{GS_mech}) of the
scalars $B^{(0)}_k$ and their dual two form $C^{(2)}_k$, we obtain the
Green-Schwarz diagram depicted in figure~2, similarly
for the dual pair $C^{(0)}_k$ and $B^{(2)}_k$. These diagrams
have the correct form to cancel the mixed anomalies~(\ref{mix_anom_1})
and ~(\ref{mix_anom_2}).

\section{NSNS tadpoles}
\label{NSNStadpoles}

Apart from the RR tadpoles considered in
section~\ref{section_RR_tadpoles}, non-supersymmetric theories
generically involve also NSNS tadpoles. In this section, we will
follow the discussion of~\cite{Blumenhagen:2001te} in computing the
NSNS tadpoles and deriving the effective scalar potential for the
closed string moduli. The analysis will be performed at next to
leading order in string perturbation theory, i.e. at open string
tree level $e^{-\phi}$.  

The massless NSNS sector fields of our model are the four-dimensional
dilaton as well as the internal metric and NSNS two form flux
moduli. In our factorized ansatz on $T^2\times T^4/\Z_3$, the moduli
of $T_1$ are the two radions $R_1$ and $R_2$ and the two form flux $b$.
In addition, $K3$ has 80 moduli. In the orbifold limit $T^4/\Z_3$,
these moduli are provided for by 11 hyper- and nine tensormultiplets where each of the nine orbifold fixed points contributes
one hyper- and one tensormultiplet. The remaining two hypermultiplets
originate from the untwisted closed string
sector~\cite{Gimon:1996ay}. The twisted NSNS moduli at each fixed
point group into a triplet state associated to the complex structure
and K\"ahler deformations of the manifold and a singlet state which
originates from the Kaluza Klein reduction of the ten dimensional
$\Omega$ odd form $B^{(2)}$ on $\Sigma_k$. The NSNS triplet and the RR
scalar provide the bosonic degrees of freedom of a hypermultiplet, and
the NSNS scalar together with the RR two form belong to a tensormultiplet at each orbifold fixed point~\cite{Douglas:1996sw}.

The computation of NSNS tadpoles is
completely analogous to the one of the RR tadpoles: The tadpoles are
extracted from the infrared divergences in the tree channel Klein
bottle, annulus and M\"obius strip amplitude. These three
contributions lead to a sum of perfect squares which can be identified
with the disc tadpoles of the various NSNS moduli of the
theory. In the $\OR_1$ orientifold model on $T^2 \times T^4/\Z_3$,
three contributions to the tadpoles arise at next to leading
order. Two of them, the dilaton tadpole and the tadpole of the complex
structure on $T_1$, originate from the untwisted part of the
amplitudes. They have the interpretation given
in~\cite{Blumenhagen:2001te} which we will shortly repeat
here. Additionally, a third tadpole is
generated by the twisted moduli corresponding the fixed points of $T^4/\Z_3$.

In detail, for the dilaton tadpole we obtain
\begin{equation}\label{dilaton_tadpole}
\langle \phi \rangle_D=\frac{1}{\sqrt{\text{Vol}(T^6)}}
 \left(\sum_{a=1}^K N_a\text{Vol}(\text{D}8_a)-16\text{Vol}(\text{O}8_a)\right)
\end{equation}
with
\begin{align}
\text{Vol}(\text{D}8_a)&=\omega L_a=\omega \sqrt{(n_aR_1)^2+((m_a+bn_a)R_2)^2},\nonumber\\
\text{Vol}(\text{O}8_a)&=\omega R_1,\nonumber
\end{align}
and the tadpole for the imaginary part of the complex structure on
$T_1$ is given by
\begin{equation}\label{com_structure_tadpole} 
\langle u \rangle_D=\frac{1}{\sqrt{\text{Vol}(T^6)}}
 \left(\sum_{a=1}^K N_a\frac{(n_aR_1)^2-((m_a+bn_a)R_2)}{L_a}
-16\text{Vol}(\text{O}8_a)\right).
\end{equation}
As explained in~\cite{Blumenhagen:2001te}, the dilaton tadpole gives
the effective tension of the brane configuration in four
dimensions. In contrast to the type IIB models constructed
in~\cite{Aldazabal:2001dg,Aldazabal:2001cn}, the real part of the
complex structure in the T-dual picture with background fields, i.e. the antisymmetric NSNS two form, is not a modulus
of the orientifold theory, and therefore we only obtain a tadpole for
the imaginary part. Defining
$u=\sqrt{\left|U_2\right|}=\sqrt{R_1/R_2}$, the dilaton and the
complex structure tadpole can be cast in the form
\begin{align}\label{dilaton_tadpole_u}
\langle \phi \rangle_D &= \sqrt{\omega}\left(\sum_{a=1}^K
  N_a{\cal L}_a-16u\right)\\ 
\label{com_structure_tadpole_u} 
\langle u \rangle_D &= u\frac{\partial}{\partial u} \left(
\sqrt{\omega} \left(\sum_{a=1}^K
  N_a{\cal L}_a-16u\right)\right)
\end{align}
with
\begin{equation}
{\cal L}_a(U)=\sqrt{(n_au)^2+((m_a+U_1n_a)\frac{1}{u})^2}.\nonumber
\end{equation}
The formulas (\ref{dilaton_tadpole_u}) and
(\ref{com_structure_tadpole_u}) reflect the fact that, regarding $T_1$
where the  reflexion ${\cal R}_1$ acts, only the
left-right symmetric states of the closed string Hilbert space,
in this case the complex structure modulus, couple to the crosscaps
and boundary states whereas the left-right antisymmetric ones,
here the K\"ahler modulus, do not. In addition, we expect to find couplings to
the moduli 
of $K3$. Indeed, a third tadpole arises from the twisted sector which
can be cast into the form
\begin{equation}
\langle \varphi_k \rangle_D=\left(\sum_a \tr(\gamma_a){\cal L}_a-4u\right).
\end{equation}
From
\begin{equation}
\langle \phi \rangle_D\sim\frac{\partial V}{\partial \phi} 
\qquad \langle u \rangle_D\sim\frac{\partial V}{\partial u}
\qquad \langle \varphi_k \rangle_D\sim\frac{\partial V}{\partial \varphi_k}
\end{equation}
we can derive an ansatz for the scalar potential in the string frame
of the form
\begin{equation}\label{NSNS_potential}
V(\phi,U,\varphi_k)=e^{-\phi}\left(\sum_{a=1}^K N_a{\cal L}_a-16u
+\varphi_k \left(\sum_{a=1}^K \tr(\gamma_a){\cal L}_a-4u \right)\right).
\end{equation}
\newline
This potential is only leading order in string theory although higher
powers of the complex structure modulus occur. 
The ansatz (\ref{NSNS_potential}) for the scalar potential can be
compared with the field theory expectation obtained from the
Dirac-Born-Infeld action of a D$9_a$ brane with constant magnetic and
electric background flux in the T-dual picture in the limit
$\varphi_k\rightarrow 0$,
\begin{equation}
S_{BI}=-T_9\int_{D9_a}\dd^{10}x e^{-\phi}\sqrt{\text{det}\left(G+{\cal F}_a\right)} 
\end{equation}
with the D9-brane tension
$T_9=\frac{\sqrt{\pi}}{16\kappa_0}4\pi^2\alpha'$ and the constant
values on $T_1$
\begin{equation}
G=\unity_2, \qquad 
{\cal F}^{45}_a=\left(B+F\right)^{45}_a
=\frac{(m_a+bn_a)\alpha^{\prime}}{n_aR_1R_2}. 
\end{equation}
In addition, to lowest order in the $K3$ moduli the relation
\begin{equation}
\text{det}G(K3)=\text{vol}(K3)=\omega
\end{equation}
is valid. The dependence on the blow-up modes $\varphi_k$ in the orbifold limit
$T^4/\Z_3$ seems to be much more complicated and will not be further
pursued here.

The scalar potential (\ref{NSNS_potential}) computed from string
theory is unstable to lowest order. This means that the minimum of the
theory is not chosen in an appropriate way and hints to an instability
of the brane configuration. In the T-dual theory, the tilting of branes
towards the $x^4$ axis corresponds to the dynamical decompactification
to the six-dimensional supersymmetric theory.
As mentioned in section
\ref{section_chiral_spectrum}, it seems to be impossible to obtain a
consistent chiral theory which does not contain any tachyon at
all. The problem of stability in the context of tachyons in a purely
toroidal compactification has also been addressed
in~\cite{Rabadan:2001mt}.

\section{Examples}
\label{examples}

In this section we discuss four models in view of their  phenomenological
relevance. The tadpole conditions~(\ref{tad_con_N1}), (\ref{tad_con_N2}) severely
restrict the possible choices of gauge groups. For example, the GUT
gauge group $SU(5)$ can only be obtained from $N^0_a=5$ if we restrict
our attention to branes (i.e. we do not want to include anti-branes) and we would have to
introduce at least to more stacks of branes leading to exotic
matter. Furthermore, the generic spectrum in
table~\ref{tab_gen_spect} shows that only an even number of
antisymmetric representations of $SU(N^0_a=5)$ can be
engineered. Therefore, we will not further pursue GUT models, but show
two models which include the  gauge group $SU(3)\times SU(2)\times
U(1)_Y$ and two left-right symmetric models which contain $SU(3)\times
SU(2)_L\times SU(2)_R \times U(1)_{B-L}$. In order to obtain a
phenomenologically appealing spectrum, we also include parallely displaced
branes and anti-branes. In all four models we choose the
non-trivial background $b=1/2$ as only in this case an odd number of
generations is achievable.  
\subsection{Example 1a: $SU(3)\times SU(2) \times U(1)^3$ and four generations}
\label{sec_ex1a}

\begin{figure}
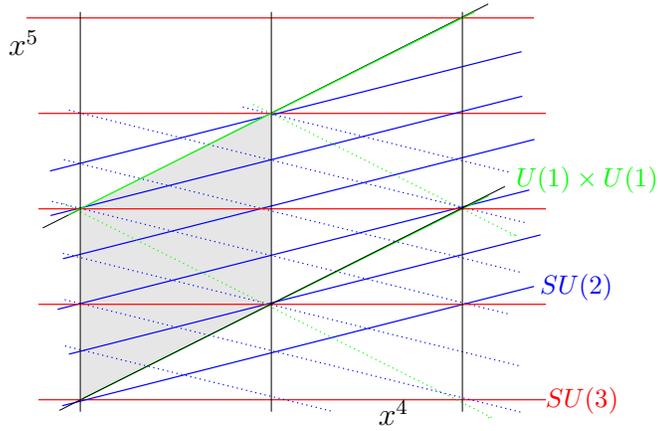
  
\begin{center}
\input brane_conf_ex1a.pstex_t
\end{center}
\caption{Example 1a: Brane Configuration on $T_1$. The shaded area
  emphasises the fundamental cell of the torus. Solid lines denote
  D-branes, dotted lines denote their mirror images.}
\label{brane_conf_1a}
\end{figure}

In the first example, we choose three different stacks of branes,
\begin{eqnarray}
N^1_A=3, \qquad (n_A,m_A)=(2,-1),\nonumber\\
N^0_B=2, \qquad (n_B,m_B)=(4,-1),\label{Set_ex1}\\
N^1_C=1, \qquad (n_C,m_C)=(1,0).\nonumber
\end{eqnarray}
The brane configuration is depicted in
figure~\ref{brane_conf_1a}. This model has already been presented in~\cite{Honecker:2001dj}.
The stack of branes of type $A$ is $\OR_1$ invariant. 
 Thus, the modified tadpole
conditions~(\ref{tad_con_N1}), (\ref{tad_con_N2}) hold and the spectrum can be
read off from tables~\ref{tab_gen_spect}
and~\ref{tab_gen_spect_OR_inv}. In this attempt, we only include
branes and require that quarks have no tachyonic partners. In
addition, we want to avoid exotic matter which would arise from
additional stacks of branes with non-abelian gauge groups.  
This fixes the numbers $N^1_A$ and $N^0_B$ as well as the corresponding 
wrapping numbers $n_A,n_B$ along the $\R_1$ invariant direction. It
also fixes the number of quark generations to be even. The spectrum
obtained from the setting~(\ref{Set_ex1}) is displayed in
table~\ref{tab_ex1a} where we have also listed the original $(Q^i_a)$ and
anomaly-free $(Q_Y, \Tilde{Q})$ $U(1)$ charges. The factor $U(1)_{1,A}$
 which arises from the $\OR_1$ stack of branes is anomaly-free by
itself. In addition, there are two more anomaly-free linear combinations, 
\begin{eqnarray}
Q_Y=\frac{Q^1_A}{3}+ Q_C^1-Q_C^2,\\
\Tilde{Q}=\frac{Q^0_B}{4}+Q_C^1+Q_C^2,\nonumber
\end{eqnarray}
where $Q_Y$ can be interpreted as hypercharge for the left- and
right-handed quarks and leptons. The remaining anomalous $U(1)$
factor acquires a mass by the generalized Green-Schwarz mechanism as
described in 
section~\ref{sec_mixed_an} and decouples from the effective theory.   
\renewcommand{\arraystretch}{1.5}
\begin{table}[ht]
  \begin{center}
    \begin{equation*}
      \begin{array}{|c||c||c||c|c|c||c|c|c|} \hline
        \multicolumn{9}{|c|}{\rule[-3mm]{0mm}{8mm} \text{\bf
         Chiral fermionic spectrum for example 1a}} \\ \hline\hline
        & \text{mult.} &\text{rep. of } SU(3) \times SU(2) 
        & Q_C^1 & Q_C^2 & Q^0_B
        & Q^1_A & Q_Y
        & \Tilde{Q} 
\\ \hline\hline
AB \alpha^1 & 2 & (\Bar{3},2) & 0& 0& -1 & -1 & -1/3&-1/4\\ 
 \alpha^2 & 2 & (3,2) & 0& 0& -1 & 1 & 1/3&-1/4\\\hline 
AC \alpha^0 & 2 & (\Bar{3},1) & 1 &0 & 0& -1 & 2/3&1\\
& 2 & (3,1) & 0 & 1 &0 & 1 & -2/3&1\\ 
 \alpha^1 & 1 & (3,1) &  0& -1 &  0& 1 & 4/3&-1\\ 
 \alpha^2 & 1 & (\Bar{3},1) &-1& 0& 0& -1 & -4/3&-1\\\hline
BC \alpha^1 & 1 & (1,2) &-1& 0& 1 &  0& -1&-3/4\\
 \alpha^2 & 1 & (1,2) & 0&-1&1&  0& 1&-3/4\\\hline
BC'\alpha^1 & 3 & (1,2) &-1 & 0& -1 &  0& -1&-5/4\\
\alpha^2 & 3 & (1,2) &0&-1&-1&  0& 1&-5/4\\\hline
BB'\alpha^0 & 4 & (1,1_a) &0&0&2&0 &0&1/2\\
& 6 & (1,1_a)+(1,3_s) &0&0&2&0 &0&1/2\\\hline
CC'\alpha^0 & 2 & (1,1) &1&1&0&0 &0 &2 
\\ \hline
      \end{array}
    \end{equation*}
  \end{center}
\caption{Chiral fermionic spectrum for example 1a}
\label{tab_ex1a}
\end{table}
In the $AC\alpha^0$, $BB'\alpha^0$ and $CC'\alpha^0$ sectors,
tachyonic pseudo-superpartners occur, whereas all other sectors have
either massless or massive scalar partners transforming in the same
representation.


\subsection{Example 1b: $SU(3)\times SU(2) \times U(1)^2$ and four generations}
\label{sec_ex1b}

\begin{figure}
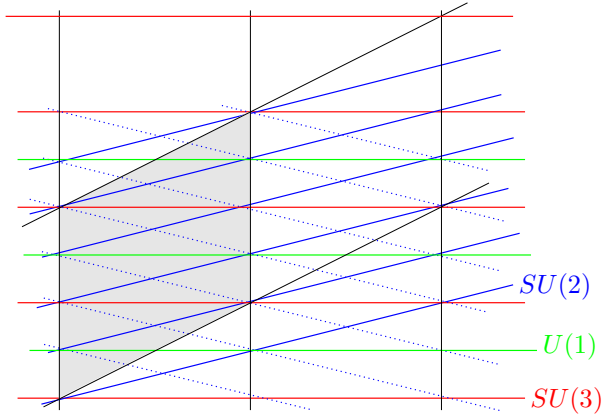
  
\begin{center}
\input brane_conf_ex1b.pstex_t
\end{center}
\caption{Example 1b: Brane Configuration on $T_1$}
\label{brane_conf_1b}
\end{figure}

The chiral fermion content of example 1a discussed in
section~\ref{sec_ex1a} contains a different number of particles and
anti-particles , namely four candidates for quarks and six
candidates for anti-quarks, and also a different amount of quarks and
leptons. Bearing in mind the considerations made in engineering model
1a, we modify the third type of brane $C$ such that the amount of
quarks and leptons matches. This can be achieved by 
\begin{eqnarray}
N^1_A=3, \qquad (n_A,m_A)=(2,-1),\nonumber\\
N_B^0=2, \qquad (n_B,m_B)=(4,-1),\\
N_C^1=1, \qquad (n_C,m_C)=(2,-1),\nonumber
\end{eqnarray}
where the stacks $C$ and $A$ are parallely displaced. The distance of
the branes serves to break $SU(4)$ down to $SU(3)\times U(1)$. In the
T-dual picture, distances translate into Wilson lines. The
brane configuration is displayed in figure~\ref{brane_conf_1b}. As one
can easily see from this figure, locating the stack $C$ at $x^5=R_2/4$
and taking into account lattice shifts gives again an $\OR_1$
configuration.\footnote{Locating a brane $c$ at $x^5=R_2/4$ is
  convenient, but not necessary. For $m_c+bn_c=0$,
  equation~(\ref{tad_con_T_1}) does not give any constraint on the
  $\gamma$ matrices. The second choice consistent with the closure of
  the orbifold group is $\gamma^c_{\OR_1}=\gamma^{c'}_{\OR_1}=\unity_{N_c}$ and
  $\gamma^c_{\Theta}=\gamma^{-1,c'}_{\Theta}=\text{diag}\left(\unity_{N_c^0}, e^{2\pi i/3}_{N_c^1}, 
e^{-2\pi i/3}_{N_c^2}\right)$ for $c\neq c'$. In this case, $N_c^1$ and $N_c^2$ can
be chosen independently.}
In this case, we obtain four generations of quarks an leptons as well
as several exotic fermions. The complete spectrum is listed in
table~\ref{tab_ex1b}. In this case, $Q_B^0$ becomes massive while
$Q_A^1$ and $Q_C^1$ are anomaly-free by themselves. The linear
combination $Q_Y=\frac{Q_A^1}{3}+Q_C^1$ can be interpreted as the SM
hypercharge. 
\renewcommand{\arraystretch}{1.5}
\begin{table}[ht]
  \begin{center}
    \begin{equation*}
      \begin{array}{|c||c||c||c||c|c|} \hline
        \multicolumn{6}{|c|}{\rule[-3mm]{0mm}{8mm} \text{\bf
         Chiral fermionic spectrum for example 1b}} \\ \hline\hline
        & \text{mult.} &\text{rep. of } SU(3) \times SU(2) 
        & Q_B^0 & Q_A^1 & Q_C^1
\\ \hline\hline
AB \alpha^1 & 2 & (\Bar{3},2) &-1& -1 &0  \\ 
\alpha^2 & 2 & (3,2) &-1& 1 & 0 \\\hline 
BC \alpha^1 & 2 & (1,2) &-1&  0 & -1\\
\alpha^2 & 2 & (1,2) &-1& 0  & 1\\\hline
BB'\alpha^0 & 10 & (1,1_a) &2& 0 &0\\
&  6 & (1,3_s) &2& 0 &0
\\ \hline
      \end{array}
    \end{equation*}
  \end{center}
\caption{Chiral fermionic spectrum for example 1b}
\label{tab_ex1b}
\end{table}


\subsection{Example 2a: $SU(3)\times SU(2)_L\times SU(2)_R
  \times  SO(8) \times U(1)^3$ and three generations}
\label{sec_ex2a}

\begin{figure}
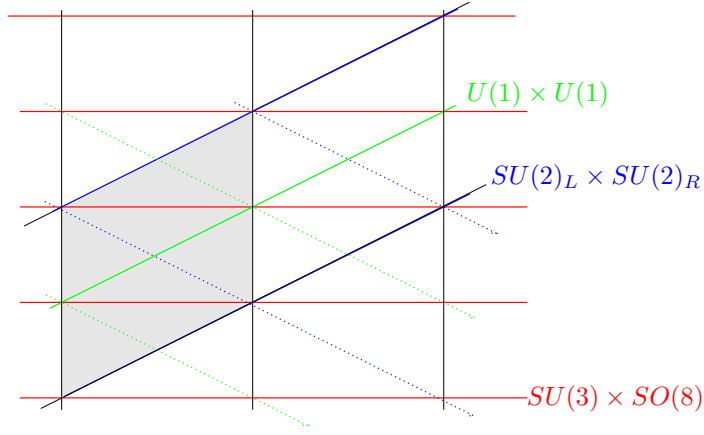
  
\begin{center}
\input brane_conf_ex2a.pstex_t
\end{center}
\caption{Example 2a: Brane Configuration on $T_1$}
\label{brane_conf_2a}
\end{figure}

So far, we have only managed to engineer an even number of generations
of the SM gauge group even though we have switched on a non-trivial
background field $b$. The following examples are chosen to be left-right
symmetric and contain three generations of left-handed quarks and
leptons. We again choose the $SU(3)$
factor to arise from the $\OR_1$ invariant position and the
$SU(2)_L\times SU(2)_R$ factors to be supported by branes at non-trivial
angles. In order to fulfill the tadpole
conditions~(\ref{tad_con_N1}), (\ref{tad_con_N2}), an additional gauge group
$SO(8)$ as well as an anti-brane have to be included. The brane
configuration of our first choice 
\begin{eqnarray}
\begin{array}{l}\Bigl.
N^0_A=8\\
N^1_A=3 \end{array}\Bigr\} \qquad (n_A,m_A)=(2,-1),\nonumber\\
N_B^1=2, \qquad (n_B,m_B)=(1,0),\\
N_C^1=1, \qquad (n_C,m_C)=(-1,0),\nonumber
\end{eqnarray}
with a parallel displacement of the branes $B$ and anti-brane $C$ is
shown in figure~\ref{brane_conf_2a}.
The complete spectrum is listed in table~\ref{tab_ex2a} of
appendix~\ref{app_spectra_2a2b}. It contains three generations of
quarks and leptons as well as their anti-particles. In addition, it
contains exotic matter transforming in the fundamental representation
of $SO(8)$, a $(2,2)$ of $SU(2)_L\times SU(2)_R$ whose tachyonic
partner could be interpreted as a non-standard Higgs particle and
several singlets of the non-abelian gauge groups. The anomaly-free
$U(1)$s are given by 
\begin{eqnarray}
Q_{B-L}=-\frac{1}{3}Q^1_A+Q^1_C-Q^2_C,\nonumber\\
Q^{\prime}=-\frac{2}{3}Q^1_A+Q^1_B-Q^2_B,\\
Q^{\prime\prime}=\frac{1}{4}\left(Q^1_B+Q^2_B+2Q^1_C+2Q^2_C\right),\nonumber
\end{eqnarray}
where $Q_{B-L}$ can be interpreted as Baryon - Lepton number occuring
in left-right symmetric models. 

There are two facts which have to be taken care of when including anti-branes. On
the one hand, the GSO-projection in the brane - anti-brane sector is
opposite to the usual one and results in selecting the reverse
chirality. On the other hand, the $\OR_1$ projection in the $CC'$
sector selects the symmetric instead of the antisymmetric
representation in the R sector.  
Due to the displacement of the stacks $B$ and $C$, there will
be no tachyons stretched between parallel branes and anti-branes as long as
$R_1$ and $R_2$ are chosen big enough. 

In this example, tachyonic pseudo-superpartners
$\psi^1_{\Delta\varphi-1/2}|0\rangle_{NSNS}$ occur in the $AB\alpha^0$,
 $BB'\alpha^0$ and $CC'\alpha^0$ sectors.
In the $AC\alpha^0$ and $BC'\alpha^0$ sectors, the reversed
GSO-projection leaves the tachyonic groundstate $|0\rangle_{NSNS}$
invariant.


\subsection{Example 2b: $SU(3)\times SU(2)_L\times SU(2)_R
  \times  SO(8) \times U(1)^2$ and three generations}
\label{sec_ex2b}

\begin{figure}
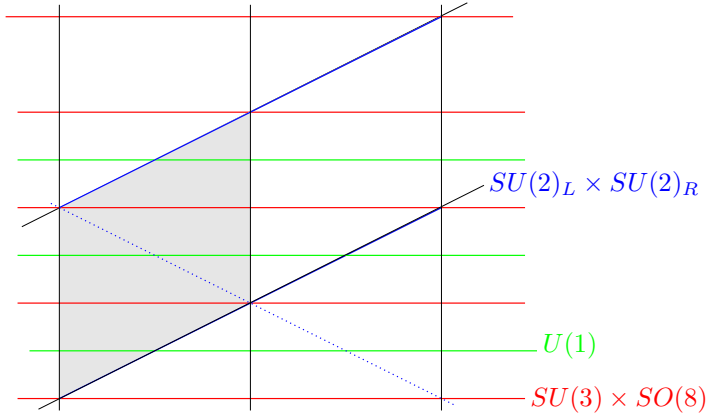
  
\begin{center}
\input brane_conf_ex2b.pstex_t
\end{center}
\caption{Example 2b: Brane Configuration on $T_1$}
\label{brane_conf_2b}
\end{figure}

As a last example, we start with the same $SU(3)\times SU(2)_L\times
SU(2)_R$ configuration as in example 2a, but choose the anti-brane $C$
to be $\OR_1$ invariant and parallely displaced w.r.t. the $SU(3)$
stack. The brane positions resulting from 
\begin{eqnarray}
\begin{array}{l}\Bigl.
N^0_A=8\\
N^1_A=3 \end{array}\Bigr\} \qquad (n_A,m_A)=(2,-1),\nonumber\\
N_B^1=2, \qquad (n_B,m_B)=(1,0),\\
N_C^1=1, \qquad (n_C,m_C)=(-2,1)\nonumber
\end{eqnarray}
are displayed in figure~\ref{brane_conf_2b}.
The complete chiral spectrum is listed in table~\ref{tab_ex2b} of
appendix~\ref{app_spectra_2a2b}, and the anomaly free $U(1)$ factors
are given by
\begin{eqnarray}
 Q_{B-L}=-\frac{1}{3}Q^1_A-Q^1_C,\\
 Q^{\prime} =Q^1_B-Q^2_B+2Q^1_C.\nonumber
\end{eqnarray}
In this case, the spectrum contains three generations of left- and
right-handed quarks and
leptons beside some exotic matter, the GSO-projection is reversed in
the $AC$ and $BC$ sectors, and tachyons with the same representation of
the gauge group as the fermions appear in the $AB\alpha^0$, $BB'\alpha^0$ and
$BC\alpha^0$ sectors. 


\section{Discussion and Conclusions}
\label{discussion}

In this paper, we have derived the tadpole conditions and chiral spectra of
type IIA orientifolds on $T^2\times T^4/\Z_3$ with D8-branes at
non-trivial angles on $T^2$. This class of models is T-dual to
$\OR_1I_4$ orientifolds with the same orbifold group $\Z_3$ (where
$I_4$ is the reflexion of all four coordinates of the orbifold) 
but
D4-branes instead of D8-branes. The dual D4-branes are transverse to the
four dimensional orbifold. As the Planck scale in these models obtained
from dimensional reduction depends on the volume $\omega$ of the orbifold, 
\begin{equation}
M_P\sim \frac{\sqrt{R_1R_2\omega}}{\lambda_s \alpha^{\prime 2}},
\end{equation}
choosing $\omega$ large can lower the string scale down to the TeV
range. In the same T-dual picture, the gauge couplings arising from D$4_a$
branes are at tree level given by~\cite{Aldazabal:2001cn} 
\begin{equation}
\frac{2\pi}{g_a^2}=\frac{M_s}{\lambda_s}L_a. 
\end{equation}
Applying this relationship to the examples discussed in the previous
section, we obtain\newline
\mbox{$\frac{\alpha_{QCD}}{\alpha_2}=2\sqrt{1+\frac{1}{16}\frac{1}{u^4}}$}
for examples 1a and 1b and
\mbox{$\frac{\alpha_{QCD}}{\alpha_2}=\frac{1}{2}\sqrt{1+\frac{1}{4}\frac{1}{u^4}}$}
for examples 2a and 2b, where $u=\sqrt{R_1/R_2}$ is as defined in section~\ref{NSNStadpoles}.
These values are only valid at tree level at the string scale
$M_s$. In order to make contact with the observed data at the
electroweak scale, the running of couplings as well as loop
corrections which might be large would have to be taken into account. 

The qualitative behaviour of Yukawa couplings~(\ref{yukawa}) can be
nicely read off from figure~\ref{brane_conf_1a} for example 1a. The
sizes $A_{ijk}$ of the smallest triangular worldsheets in units of
$\frac{R_1R_2}{\alpha'}$ are $\frac{1}{48}, \frac{1}{16},
\frac{1}{12}$ and $\frac{1}{4}$. 
There exist, however, also trilinear couplings which arise from one
single intersection point. The reason for this is that in example 1a,
two quark generations $Q^{1,2}_L$ are realised as $(\overline{3},2)$
and the other two $Q^{3,4}_L$ as $(3,2)$ in the $AB$ sector. Couplings
with higgs scalars $h$ from the $BB'$ sector are allowed by regarding the quantum
numbers. Since the position of
branes $A$ is chosen to be $\OR_1$ invariant, the intersection points
of $AB$ are also intersection points of $BB'$. 

Let us now briefly comment on example 2b. In this case, all left
handed quarks $Q^i_L$ are realised as $(\overline{3},2_L)$ while all
right handed quarks $Q^j_R$ transform as $(3,2_R)$. All quarks arise
from the $AB$ sector where $A$ is the $\OR_1$ invariant stack of branes. 
The $BB'$ sector can provide for higgs scalars $h$ in the
$(2_L,2_R)$ with $U(1)$ charges $Q^1_B=Q^2_B=\pm 1$.
The quantum numbers thus allow for trilinear couplings of the form
$hQ^i_LQ^j_R$ (for $i,j=1,2$ and $i=j=3$ since the third generation
differs in the quantum numbers $Q^1_B, Q^2_B$ from the
other two). In the same spirit, trilinear couplings $hL^i_LL^j_R$
of a higgs particle with two leptons $L^i_L$, $L^j_R$ are allowed for
$i,j=1,2$ and $i=j=3$. But in contrast to the couplings involving
quarks, the leptons arise from the $BC$ sector which does not have
any common intersection point with the $BB'$ sector. Naively, one can
therefore speculate that quark and lepton masses are generated from couplings
to the same higgs scalars $h$ acquiring a vev, and that there will be
a hierarchy of quark and lepton masses since the relevant worldsheets
are of the order $A_{hQQ}=0$ and 
$A_{hLL}\sim {\cal O}(\frac{R_1R_2}{\alpha'})$. This naive
interpretation, however, has to be handled with care since 
not all types of couplings to higgses might occur (e.g. if only  one
type of scalar particles $h$ with $Q^1_B=Q^2_B=1$ exists and no
couplings $\overline{h}QQ$ are allowed).  

\vspace{1cm}

In summary, we have computed the tadpole cancellation conditions and
generic spectra for orientifold models with intersecting D8-branes  which
have a T-dual description in terms of D4-branes at angles. The
conditions~(\ref{tad_con_N1}), (\ref{tad_con_N2}) severely restrict the
achievable gauge groups. We have given two examples with the SM
gauge group and two further left-right symmetric examples, in
which a tachyon that necessarily arises from intersecting mirror
branes might play the role of a non-standard Higgs
particle. In addition, we have computed the NSNS tadpoles in next to leading
order. In this approximation, the scalar potential depends linearly
on the twisted moduli of the orbifold, and it would be interesting to
examine in more detail if they can contribute in stabilizing the
non-supersymmetric models with branes at angles.   

Another ansatz for improved models is to consider
more complicated orbifold groups which could project out the tachyons completely.

\vspace{1cm}
\
\noindent {\bf Acknowledgments} 
 
\noindent It is a pleasure to thank Stefan F\"orste and Ralph Schreyer
for discussions. Furthermore, I acknowledge
discussions with Ralph Blumenhagen and Boris K\"ors .  
\newline  
This work is supported by the European Commission RTN programs 
\mbox{HPRN-CT-2000-00131}, 00148 and 00152.


\begin{appendix}
\section{Computation of 1-loop diagrams}
\label{app_1-loop}

\subsection{Lattice contributions on $(T^4/Z_3)/\OR_1$}
\label{app_lattice_con_T4}

The general form of the lattice sums on $T^4/Z_3$ for one $T^2$
in the loop-channel is given by ($\rho\equiv R^2/\alpha'$)
\begin{equation}
  \lu^R [\alpha](t)\equiv\sum_{m,n\in\Z}e^{-\alpha\pi t
  (m^2+mn+n^2)/\rho}.
\end{equation}
Using the Poisson resummation formula 
\begin{equation}
\sum_{x\in\Gamma}=\frac{1}{\text{vol}(\Gamma)}\sum_{p\in\Gamma^{\ast}} 
\Tilde{f}(p)  
\end{equation}
for a $d$-dimensional lattice $\Gamma$ and its dual lattice
$\Gamma^{\ast}$ with the Fourier transform
$\Tilde{f}(p)=\int_{\mathbb{R}^d}dx e^{2\pi i x\cdot p} f(x)$ 
and defining $t=1/\kappa l$ gives the lattice sums in the
tree-channel
\begin{equation}
  \lu^R [\alpha](t)=l\frac{2\kappa}{\sqrt{3}\alpha}\rho
 \lu^{1/R}\left[\frac{4\kappa}{3\alpha}\right](l).
\end{equation}
For $T^4/Z_3$, we thus obtain
\begin{align}
\text{Klein bottle:} & \qquad \label{app_latt_kb}
\left(\lu^{R_1} \lu^{R_2}\right)[1](t) = \frac{64}{3}l^2 \omega
\left(\lu^{1/R_1}\lu^{1/R_2}\right)[16/3](l)
\\
\text{Annulus:} & \qquad \label{app_latt_an}
\left(\lu^{R_1} \lu^{R_2}\right)[2](t) = \frac{4}{3}l^2 \omega
\left(\lu^{1/R_1}\lu^{1/R_2}\right)[4/3](l)
\\
\text{M\"obius strip:} & \qquad \label{app_latt_ms}
\left(\lu^{R_1} \lu^{R_2}\right)[2](t) = \frac{64}{3}l^2 \omega
\left(\lu^{1/R_1}\lu^{1/R_2}\right)[16/3](l)
\end{align}
where $\omega\equiv \rho_1 \rho_2$ is the volume of the orbifold in
units of $\alpha'$.

\subsection{Lattice contributions on $T^2/\OR_1$}
\label{app_lattice_con_T2}

\begin{figure}
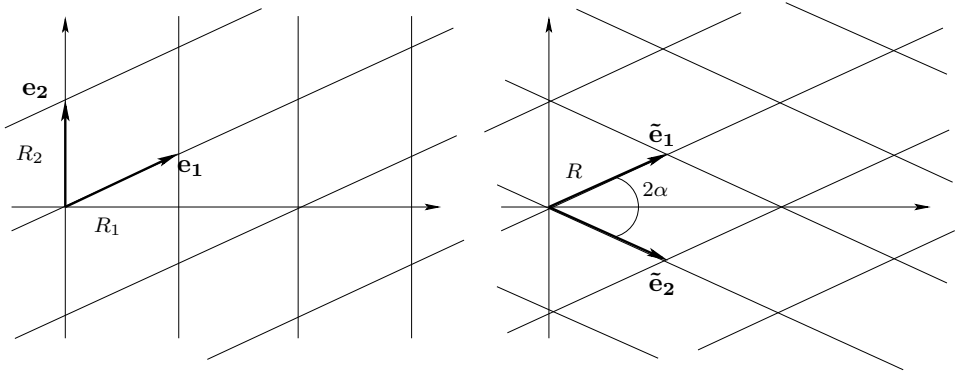
  
\begin{center}
\input lattice_comparison.pstex_t
\end{center}
\caption{Two ways to parameterize the lattice with $b=1/2$}
\label{fig_com_b-lattices}
\end{figure}

The lattice contributions of $T^2$ where the reflexion $\mathcal{R}_1$
acts are as given in~\cite{Forste:2001gb}. The result for the ${\bf b}$
type lattice can be recast in the notation
of~\cite{Blumenhagen:2001ea} with non-vanishing background $b=1/2$
field in the T-dual picture by replacing
\begin{align}
\tan{\alpha} &= \frac{R_2}{2R_1},\nonumber\\
R^2 &= R_1^2 +(R_2/2)^2,\nonumber\\
{\bf \tilde{e}_1} &= {\bf e_1},\\
{\bf \tilde{e}_2} &= {\bf e_1}-{\bf e_2},\nonumber\\
(\tilde{n},\tilde{m}) &= (n+m, -m),\nonumber  
\end{align}
where the definitions are given in figure~\ref{fig_com_b-lattices}.

\subsection{Intersection numbers and angles on the deformed $T^2$}
\label{app_intersection_numbers}

Multiplicities of chiral fermions are given in terms of the
intersection numbers where $b=0,1/2$ are the two possible choices of
background field in the T-dual picture,
\begin{align}
I_{ab} &= n_a m_b - n_b m_a,\nonumber\\
I_{aa^{\prime}} &= - \left(2m_a n_a +(2b) n_a^2 \right),\nonumber\\
I^{\OR_1}_{aa^{\prime}} &=- \left(2m_a +(2b) n_a \right),\\
I_{aa^{\prime}}-I^{\OR_1}_{aa^{\prime}} &=
-\left(2m_a +(2b) n_a\right)\left(n_a-1\right).\nonumber
\end{align}
The angles contained in the open string 1-loop amplitudes can be
re-expressed in terms of the wrapping numbers, 
\begin{align}
\frac{1}{\tan(\pi \varphi)} &= \frac{n_a}{(m_a+bn_a)}\frac{R_1}{R_2},
\nonumber\\ 
\frac{1}{\tan(\pi \Delta \varphi_{ab})} &=-\frac{1}{I_{ab}}
n_an_b\frac{R_1}{R_2}-\frac{1}{I_{ab}}(m_a+bn_a)(m_b+bn_b)\frac{R_2}{R_1},\\
\frac{I^{\OR_1}_{aa^{\prime}}}{\tan(\pi  \varphi)} 
&=- 2n_a\frac{R_1}{R_2}.\nonumber
\end{align}

\subsection{Oscillator contributions}
\label{app_osc}

Oscillator contributions to the 1-loop amplitudes can be expressed in
terms of generalized $\vartheta$ functions. The relevant formulas for
untwisted sectors without insertions can be found e.g. in~\cite{Blumenhagen:2000md}. In
addition, an insertion of $\Theta^k$ in the trace leads to
\begin{align}
\text{Klein bottle:}\qquad \label{app_osc_kb_u}
\ku^{(k)} &=  \frac{\vartheta \Bigl[\!\! \begin{array}{c} \SC 0 \\ \SC 1/2
    \end{array} \!\!\Bigr]^2}{\eta^{6}}{\displaystyle \prod_{i=2,3}}
  \frac{\vartheta \Bigl[\!\! \begin{array}{c} \SC 0 \\ \SC 1/2+2kv_i
    \end{array} \!\!\Bigr]}
  {\vartheta \Bigl[\!\! \begin{array}{c} \SC 1/2 \\ \SC 1/2+2kv_i
    \end{array} \!\!\Bigr]}(2t),
\\
\text{Annulus:}\qquad \label{app_osc_an_u}
\au_{ab}^{(k)} &=  i\frac{\vartheta \Bigl[\!\! \begin{array}{c} \SC 0 \\ \SC 1/2
    \end{array} \!\!\Bigr]}{\eta^{3}}
 \frac{\vartheta \Bigl[\!\! \begin{array}{c} \SC \Delta\varphi \\ \SC 1/2
    \end{array} \!\!\Bigr]}
  {\vartheta \Bigl[\!\! \begin{array}{c} \SC 1/2+\Delta\varphi \\ \SC 1/2
    \end{array} \!\!\Bigr]}{\displaystyle \prod_{i=2,3}}
  \frac{\vartheta \Bigl[\!\! \begin{array}{c} \SC 0 \\ \SC 1/2+kv_i
    \end{array} \!\!\Bigr]}
  {\vartheta \Bigl[\!\! \begin{array}{c} \SC 1/2 \\ \SC 1/2+kv_i
    \end{array} \!\!\Bigr]}(t),
\\
\text{M\"obius strip:}\qquad \label{app_osc_ms_u}
\msu_{a}^{(k)} &=ie^{2\pi i \varphi} 
\frac{\vartheta \Bigl[\!\! \begin{array}{c} \SC 1/2 \\ \SC 0
    \end{array} \!\!\Bigr]}{\eta^{3}}
\frac{\vartheta \Bigl[\!\! \begin{array}{c} \SC 1/2+2\varphi \\ \SC -\varphi
    \end{array} \!\!\Bigr]}
  {\vartheta \Bigl[\!\! \begin{array}{c} \SC 1/2+2\varphi \\ \SC 1/2-\varphi
    \end{array} \!\!\Bigr]}
{\displaystyle \prod_{i=2,3}}
  \frac{\vartheta \Bigl[\!\! \begin{array}{c} \SC 1/2 \\ \SC kv_i
    \end{array} \!\!\Bigr]}
  {\vartheta \Bigl[\!\! \begin{array}{c} \SC 1/2 \\ \SC 1/2+kv_i
    \end{array} \!\!\Bigr]}(t+\frac{i}{2}).
\end{align}
By modular transformation to the tree-channel, one obtains
contributions from  oscillators in the $k^{\text{th}}$ twisted sector,
\begin{align}
\text{Klein bottle:}\qquad \label{app_osc_kb_t}
\kt^{(k)} &= 
\frac{\vartheta \Bigl[\!\! \begin{array}{c} \SC 1/2 \\ \SC 0
    \end{array} \!\!\Bigr]^2}{\eta^{6}}{\displaystyle \prod_{i=2,3}}
  \frac{\vartheta \Bigl[\!\! \begin{array}{c} \SC 1/2-2kv_i \\ \SC 0
    \end{array} \!\!\Bigr]}
  {\vartheta \Bigl[\!\! \begin{array}{c} \SC 1/2-2kv_i \\ \SC 1/2
    \end{array} \!\!\Bigr]}(2l),\\
\text{Annulus:}\qquad \label{app_osc_an_t}
\at^{(k)}_{ab} &=
\frac{\vartheta \Bigl[\!\! \begin{array}{c} \SC 1/2 \\ \SC 0
    \end{array} \!\!\Bigr]}{\eta^{3}}
 \frac{\vartheta \Bigl[\!\! \begin{array}{c} \SC 1/2 \\ \SC \Delta\varphi
    \end{array} \!\!\Bigr]}
  {\vartheta \Bigl[\!\! \begin{array}{c} \SC 1/2 \\ \SC 1/2+\Delta\varphi
    \end{array} \!\!\Bigr]}
{\displaystyle \prod_{i=2,3}}
  \frac{\vartheta \Bigl[\!\! \begin{array}{c} \SC 1/2-kv_i \\ \SC 0
    \end{array} \!\!\Bigr]}
  {\vartheta \Bigl[\!\! \begin{array}{c} \SC 1/2-kv_i \\ \SC 1/2
    \end{array} \!\!\Bigr]}(2l),\\
\text{M\"obius strip:}\qquad \label{app_osc_ms_t}
\mt_{a}^{(k)} &= 
\frac{\vartheta \Bigl[\!\! \begin{array}{c} \SC 1/2 \\ \SC 0
    \end{array} \!\!\Bigr]}{\eta^{3}}
  \frac{\vartheta \Bigl[\!\! \begin{array}{c} \SC 1/2 \\ \SC \varphi
    \end{array} \!\!\Bigr]}
  {\vartheta \Bigl[\!\! \begin{array}{c} \SC 1/2 \\ \SC 1/2+\varphi
    \end{array} \!\!\Bigr]}(2l +\frac{i}{2}){\displaystyle \prod_{i=2,3}}
  \frac{\vartheta \Bigl[\!\! \begin{array}{c} \SC -kv_i \\ \SC 1/2
    \end{array} \!\!\Bigr]
\vartheta \Bigl[\!\! \begin{array}{c} \SC 1/2-kv_i \\ \SC 0
    \end{array} \!\!\Bigr]}
  {\vartheta \Bigl[\!\! \begin{array}{c} \SC 1/2-kv_i \\ \SC 1/2
    \end{array} \!\!\Bigr]
\vartheta \Bigl[\!\! \begin{array}{c} \SC -kv_i \\ \SC 0
    \end{array} \!\!\Bigr]}(4l).
\end{align}

\section{Tree channel results for $(T^2\times T^4/Z_3)/\OR_1$}
\label{app_tree}

\subsection{Crosscap states}
\label{app_tree_cross}

The crosscap conditions for the $\OR_1$-model are
\begin{eqnarray}
  \left[X^i_{L,R}(\sigma,0)-\Theta^k X^i_{R,L}(\sigma+\pi,0)\right]
  |\OR_1 \Theta^k \rangle=0,\\
  \left[X^{\overline{i}}_{L,R}(\sigma,0)-\Theta^k X^{\overline{i}}_{R,L}(\sigma+\pi,0)\right]
  |\OR_1 \Theta^k \rangle=0.
\end{eqnarray}
Inserting the mode expansion
\begin{equation}
X^i(\sigma,
\tau)=x^i+\frac{p^i_L}{2\pi}(\tau+\sigma)+\frac{p^i_R}{2\pi}(\tau-\sigma) 
+\frac{i}{2}\sum_r\frac{1}{r}\alpha^i_r e^{-ir(\tau+\sigma)}
+\frac{i}{2}\sum_s\frac{1}{s}\Tilde{\alpha}^i_s e^{-is(\tau-\sigma)}
\end{equation}
gives the following constraints on $T^4/\Z_N$
\begin{eqnarray}
\left.\begin{array}{c}
\left[p^i_L+e^{2\pi i k v_i}p^i_R
\right]\\
\left[p^{\overline{i}}_L+e^{-2\pi i k v_i}p^{\overline{i}}_R
\right]\\
\left[p^i_R+e^{2\pi i k v_i}p^i_L
\right]\\
\left[p^{\overline{i}}_R+e^{-2\pi i k v_i}p^{\overline{i}}_L
\right]\end{array}\right\}|\OR_1\Theta^k\rangle=0,
\label{app_tree_const_latt}\\
\left.\begin{array}{c}
\left[\alpha^i_r +e^{\pi i (2k v_i-r)}\Tilde{\alpha}^i_{-r}
\right]\\
\left[\alpha^{\overline{i}}_s +e^{\pi i (-2k
    v_i-s)}\Tilde{\alpha}^{\overline{i}}_{-s} 
\right]\\
\left[\Tilde{\alpha}^i_{-r} +e^{\pi i (2k v_i-r)}\alpha^i_r 
\right]\\
\left[\Tilde{\alpha}^{\overline{i}}_{-s} +e^{\pi i (-2k v_i-s)}\alpha^{\overline{i}}_s 
\right]
\end{array}\right\}|\OR_1\Theta^k\rangle=0.
\label{app_tree_const_osc}
\end{eqnarray}
The set of equations~(\ref{app_tree_const_latt}) states that for
$k=0$ windings along all four directions of the orbifold occur while
for $k \neq 0$, only Kaluza Klein momenta and windings from the first
$T^2$ contribute as discussed in appendix ${\bf B}$
of~\cite{Forste:2001gb}. The equations~(\ref{app_tree_const_osc}) are
only mutually consistent if  $r \in \Z+2kv_i$, $s
\in \Z-2kv_i$. Using the notation $n\in \Z$ , $r\in \Z (+1/2)$ for the
  R (NS) sector, the oscillator constraints can be rewritten as
\begin{eqnarray}\left.\begin{array}{c}
\left[\alpha^i_{n+2kv_i} +(-1)^n\Tilde{\alpha}^i_{-n-2kv_i}
\right]\\
\left[\alpha^{\overline{i}}_{n-2kv_i} +(-1)^n
 \Tilde{\alpha}^{\overline{i}}_{-n+2kv_i} 
\right]\end{array}\right\}|\OR_1\Theta^k\rangle=0,\\
\left.\begin{array}{c}
\left[\psi^i_{r+2kv_i} +i\eta e^{-i\pi r}\Tilde{\psi}^i_{-r-2kv_i}
\right]\\
\left[\psi^{\overline{i}}_{r-2kv_i} + i\eta e^{-i\pi r}
 \Tilde{\psi}^{\overline{i}}_{-r+2kv_i}\right] 
\end{array}\right\}|\OR_1\Theta^k\rangle=0.
\end{eqnarray}
A solution to these constraints is provided by
\begin{eqnarray}
   |\OR_1\Theta^k, \eta \rangle = & \mathcal{N}^{(k)}_C &\exp
        \Bigl\{-\sum_n\frac{(-1)^n}{n}\alpha^{\mu}_{-n} 
        \Tilde{\alpha}^{\mu}_{-n}
         -\sum_n\frac{(-1)^n}{n}\alpha^{1}_{-n}
          \Tilde{\alpha}^{1}_{-n}
         -\sum_n\frac{(-1)^n}{n}\alpha^{\Bar{1}}_{-n}
        \Tilde{\alpha}^{\Bar{1}}_{-n}\nonumber\\
         &-&\sum_{i \in \{2,3\}}\sum_n\frac{(-1)^n}{n}
         \alpha^{i}_{-n+2kv_i}
         \Tilde{\alpha}^{\overline{i}}_{-n+2kv_i}
         -\sum_{i \in \{2,3\}}\sum_n\frac{(-1)^n}{n}
         \alpha^{\overline{i}}_{-n-2kv_i}
         \Tilde{\alpha}^{i}_{-n-2kv_i}
\nonumber\\
     &-&i \eta\sum_r e^{-i\pi r}\psi^{\mu}_{-r}\Tilde{\psi}^{\mu}_{-r}
     -i \eta\sum_r e^{-i\pi r}  \psi^{1}_{-r}\Tilde{\psi}^{1}_{-r}
    -i \eta\sum_r e^{-i\pi r}\psi^{\Bar{1}}_{-r}
    \Tilde{\psi}^{\Bar{1}}_{-r}
\nonumber\\
    &-&i \eta\sum_{i \in \{2,3\}}\sum_r e^{-i\pi r}  \psi^{i}_{-r+2kv_i}
    \Tilde{\psi}^{\overline{i}}_{-r+2kv_i}
    -i \eta\sum_{i \in \{2,3\}}\sum_r e^{-i\pi r}\psi^{\Bar{i}}_{-r-2kv_i}
    \Tilde{\psi}^{i}_{-r-2kv_i}\Bigr\}|0,\eta\rangle.
\nonumber
\end{eqnarray}
The dependence on the lattice is contained in the groundstate $|0,\eta\rangle$.

\subsection{Boundary states}
\label{app_tree_boundary}
In order to reproduce the amplitudes obtained by modular
transformation from the loop channel, a boundary state at angle
$\pi\varphi$ on $T_1$ w.r.t. the $x^4$ axis
has to be of the form 
\begin{eqnarray}
   |\varphi,\Theta^k; \eta \rangle = & \mathcal{N}^{(k)}_B &\exp
        \Bigl\{-\sum_n\frac{1}{n}\alpha^{\mu}_{-n} 
        \Tilde{\alpha}^{\mu}_{-n}
         -\sum_n\frac{1}{n} e^{2\pi i \varphi}\alpha^{1}_{-n}
          \Tilde{\alpha}^{1}_{-n}
         -\sum_n\frac{1}{n}e^{-2\pi i \varphi}\alpha^{\Bar{1}}_{-n}
        \Tilde{\alpha}^{\Bar{1}}_{-n}\nonumber\\
         &-&\sum_{i \in \{2,3\}}\sum_n\frac{1}{n}
         \alpha^{i}_{-n+2kv_i}
         \Tilde{\alpha}^{\overline{i}}_{-n+2kv_i}
         -\sum_{i \in \{2,3\}}\sum_n\frac{1}{n}
         \alpha^{\overline{i}}_{-n-2kv_i}
         \Tilde{\alpha}^{i}_{-n-2kv_i}
\nonumber\\
     &-&i \eta\sum_r \psi^{\mu}_{-r}\Tilde{\psi}^{\mu}_{-r}
     -i \eta\sum_r  e^{2\pi i \varphi}\psi^{1}_{-r}\Tilde{\psi}^{1}_{-r}
    -i \eta\sum_r e^{-2\pi i \varphi}\psi^{\Bar{1}}_{-r}
    \Tilde{\psi}^{\Bar{1}}_{-r}
\nonumber\\
    &-&i \eta\sum_{i \in \{2,3\}}\sum_r \psi^{i}_{-r+2kv_i}
    \Tilde{\psi}^{\overline{i}}_{-r+2kv_i}
    -i \eta\sum_{i \in \{2,3\}}\sum_r\psi^{\Bar{i}}_{-r-2kv_i}
    \Tilde{\psi}^{i}_{-r-2kv_i}\Bigr\}|0,\eta\rangle.
\nonumber
\end{eqnarray}
As for the crosscap states, the groundstate $|0,\eta\rangle$ contains
Kaluza-Klein momentum and winding eigenvalues from $T^2$ and windings
from $T^4/\Z_N$.

\subsection{Zero modes and GSO invariant states}
\label{app_tree_GSO}

We present the following discussion for the crosscap states. The GSO
projections on boundary states are completely analogous.
 
\subsubsection{NSNS sector}
\label{app_tree_GSO_NSNS}

In the NSNS sectors, the GSO projection on the ground state is
determined by requiring  tachyonic ground states to be
unphysical. Therefore, the GSO-invariant combination is
\begin{equation}
|\OR_1\Theta^k\rangle_{NSNS} =|\OR_1\Theta^k,+\rangle_{NSNS}-
 |\OR_1\Theta^k,-\rangle_{NSNS} 
\end{equation}

\subsubsection{Untwisted RR sector}
\label{app_tree_GSO_U_RR}

Defining ($i=2,3$)
\begin{eqnarray}
\psi^{\mu}_{\eta}&=&\frac{1}{\sqrt{2}}\left(
\psi^{\mu}_0+i\eta\Tilde{\psi}^{\mu}_0\right),\nonumber\\
\psi^{1}_{\eta}&=&\frac{1}{\sqrt{2}}\left(
\psi^{1}_0+i\eta\Tilde{\psi}^{\overline{1}}_0\right), \qquad
\psi^{\overline{1}}_{\eta}=\frac{1}{\sqrt{2}}\left(
\psi^{\overline{1}}_0+i\eta\Tilde{\psi}^{1}_0\right),\\
\psi^{i}_{\eta}&=&\frac{1}{\sqrt{2}}\left(
\psi^{i}_0+i\eta\Tilde{\psi}^{i}_0\right), \qquad
\psi^{\overline{i}}_{\eta}=\frac{1}{\sqrt{2}}\left(
\psi^{\overline{i}}_0+i\eta\Tilde{\psi}^{\overline{i}}_0\right),\nonumber
\end{eqnarray}
the non-trivial commutation relations are
\begin{eqnarray}
\left\{\psi^{\mu}_+, \psi^{\mu}_-\right\}&=&1,\nonumber\\
\left\{\psi^{1}_+, \psi^{\overline{1}}_-\right\}
&=&\left\{\psi^{1}_-, \psi^{\overline{1}}_+\right\}=1,\\
\left\{\psi^{i}_+, \psi^{\overline{i}}_-\right\}
&=&\left\{\psi^{i}_-, \psi^{\overline{i}}_+\right\}=1.\nonumber
\end{eqnarray}
The crosscap conditions from the zero modes in the RR-sector on the
groundstate then read
\begin{equation}
\left. \begin{array}{c}
\psi^{\mu}_{\eta}\\
\psi^{1}_{\eta}\\
\psi^{\overline{1}}_{\eta}\\
\psi^{i}_{\eta}\\
\psi^{\overline{i}}_{\eta}
\end{array}\right\}|\OR_1,\eta\rangle_{_{RR}}^0=0,
\end{equation}
and the zero mode parts of the GSO projections are given by
\begin{align}
(-1)^{F}=\prod_{m=2}^9 &\sqrt{2}\psi^m_0\\
=\prod_{\mu=2,3}&\left(\psi^{\mu}_{+}+\psi^{\mu}_{-}\right)
\cdot
\frac{1}{2i}\left(\psi^{1}_{+}+\psi^{1}_{-}
+\psi^{\overline{1}}_{+}+\psi^{\overline{1}}_{-}\right)
\left(\psi^{1}_{+}+\psi^{1}_{-}
-\psi^{\overline{1}}_{+}-\psi^{\overline{1}}_{-}\right)\nonumber\\
\prod_{i=2,3}&\frac{1}{2i}
\left(\psi^{i}_{+}+\psi^{i}_{-}
+\psi^{\overline{i}}_{+}+\psi^{\overline{i}}_{-}\right)
\left(\psi^{i}_{+}+\psi^{i}_{-}
-\psi^{\overline{i}}_{+}-\psi^{\overline{i}}_{-}\right)\nonumber\\
(-1)^{\Tilde{F}}=\prod_{m=2}^9 &\sqrt{2}\Tilde{\psi}^m_0\\
=\prod_{\mu=2,3}&\frac{1}{i}\left(\psi^{\mu}_{+}-\psi^{\mu}_{-}\right)
\cdot
\frac{1}{2i}\left(\psi^{1}_{+}-\psi^{1}_{-}
+\psi^{\overline{1}}_{+}-\psi^{\overline{1}}_{-}\right)
\left(\psi^{1}_{+}-\psi^{1}_{-}
-\psi^{\overline{1}}_{+}+\psi^{\overline{1}}_{-}\right)\nonumber\\
\prod_{i=2,3}&\frac{-1}{2i}
\left(\psi^{i}_{+}-\psi^{i}_{-}
+\psi^{\overline{i}}_{+}-\psi^{\overline{i}}_{-}\right)
\left(\psi^{i}_{+}-\psi^{i}_{-}
-\psi^{\overline{i}}_{+}+\psi^{\overline{i}}_{-}\right).\nonumber
\end{align}
Defining
\begin{equation}
|\OR_1,-\rangle^0 \equiv \left[
\left(\prod_{\mu=2,3}\psi^{\mu}_{-}\right)
\left(\psi^{1}_{-}\psi^{\overline{1}}_{-}\right)
\left(\prod_{i=2,3}\psi^{i}_{-}\psi^{\overline{i}}_{-}\right)
\right]|\OR_1,+\rangle^0,
\end{equation}
the action of the complete GSO-projector can be rephrased as
\begin{eqnarray}
(-1)^{F}|\OR_1,+\rangle =-(-1)^{\Tilde{F}}|\OR_1,+\rangle=-i|\OR_1,-\rangle,\\
(-1)^{F}|\OR_1,-\rangle =-(-1)^{\Tilde{F}}|\OR_1,-\rangle =i|\OR_1,+\rangle,
\end{eqnarray}
and
\begin{equation}
|\OR_1,+\rangle_{_{RR}}-i|\OR_1,-\rangle_{_{RR}} \qquad
\text{ is invariant w.r.t.} \qquad 
P_{GSO}=\frac{1+(-1)^F}{2}\frac{1-(-1)^{\Tilde{F}}}{2}.
\end{equation}

\subsubsection{Twisted RR sectors}
\label{app_tree_GSO_T_RR}

For $k\neq 0$, the zero mode conditions read 
\begin{equation}
\left. \begin{array}{c}
\psi^{\mu}_{\eta}\\
\psi^{1}_{\eta}\\
\psi^{\overline{1}}_{\eta}\\
\end{array}\right\}|\OR_1\Theta^k,\eta\rangle_{_{RR}}^0=0.
\end{equation}
The zero mode parts of the GSO projection operators are now given by
\begin{align}
(-1)^{F}=\prod_{m=2}^5 & \sqrt{2}\psi^m_0\\
=\prod_{\mu=2,3}&\left(\psi^{\mu}_{+}+\psi^{\mu}_{-}\right)
\cdot
\frac{1}{2i}\left(\psi^{1}_{+}+\psi^{1}_{-}
+\psi^{\overline{1}}_{+}+\psi^{\overline{1}}_{-}\right)
\left(\psi^{1}_{+}+\psi^{1}_{-}
-\psi^{\overline{1}}_{+}-\psi^{\overline{1}}_{-}\right),\nonumber\\
(-1)^{\Tilde{F}}=\prod_{m=2}^5 &\sqrt{2}\Tilde{\psi}^m_0\\
=\prod_{\mu=2,3} & \frac{1}{i}\left(\psi^{\mu}_{+}-\psi^{\mu}_{-}\right)
\cdot
\frac{1}{2i}\left(\psi^{1}_{+}-\psi^{1}_{-}
+\psi^{\overline{1}}_{+}-\psi^{\overline{1}}_{-}\right)
\left(\psi^{1}_{+}-\psi^{1}_{-}
-\psi^{\overline{1}}_{+}+\psi^{\overline{1}}_{-}\right).\nonumber
\end{align}
Using
\begin{equation}
|\OR_1\Theta^k,-\rangle^0 \equiv \left[
\left(\prod_{\mu=2,3}\psi^{\mu}_{-}\right)
\left(\psi^{1}_{-}\psi^{\overline{1}}_{-}\right)
\right]|\OR_1\Theta^k,+\rangle^0
\end{equation}
leads to the action of the zero mode part of the GSO projector on the
groundstates
\begin{eqnarray}
(-1)^{F}|\OR_1\Theta^k,+\rangle^0
=-(-1)^{\Tilde{F}}|\OR_1\Theta^k,+\rangle^0
=i|\OR_1\Theta^k,-\rangle^0, \\  
(-1)^{F}|\OR_1\Theta^k,-\rangle^0
=-(-1)^{\Tilde{F}}|\OR_1\Theta^k,-\rangle^0=-i|\OR_1\Theta^k,+\rangle^0.  
\end{eqnarray}
These relations carry over to the excited states. Thus,
\begin{equation}
|\OR_1\Theta^k,+\rangle_{_{RR}}+i|\OR_1\Theta^k,-\rangle_{_{RR}} \qquad
\text{ is invariant w.r.t.} \qquad 
P_{GSO}=\frac{1+(-1)^F}{2}\frac{1-(-1)^{\Tilde{F}}}{2}.
\end{equation}

\section{Massless states and chiral fermions for $T^2\times T^4/\Z_3$}
\label{app_massless_spectrum}

The lightest mass eigenstates are distinguished by their $\Theta$
eigenvalues. Defining $\alpha\equiv e^{2\pi i/3}$, the lightest
bosonic and fermionic states between branes $a$ and $b$ at angle $\pi \Delta
\varphi$ on $T^2$ are listed in the following tables.

\renewcommand{\arraystretch}{1.3}
    \begin{equation*}
  \begin{array}{cc}
      \begin{array}{|c||l|c|c|} \hline
        \multicolumn{4}{|c|}{\rule[-3mm]{0mm}{8mm} \text{\bf
            Bosonic open spectrum of $T^2 \times T^4/\Z_3$}} \\ \hline\hline
        \text{on }T^2 & \text{state} &\text{mass}& \Z_3\\ \hline\hline
        \Delta\varphi=0  &
        \psi^{\mu}_{-1/2}|0\rangle&0
        & 1 \\ \hline
        &\psi^{1,\Bar{1}}_{-1/2}|0\rangle&0
        & 1 \\ \hline
        &\psi^{2,\Bar{3}}_{-1/2}|0\rangle&0
        &\alpha\\ \hline
        &\psi^{\Bar{2},3}_{-1/2}|0\rangle&0
        &\alpha^2\\
        \hline\hline 
\Delta\varphi \neq 0       
        &\psi^{\mu}_{-1/2}|0\rangle &\frac{1}{2}\Delta\varphi 
        & 1 \\ \hline
        & \psi^{1}_{\Delta\varphi-1/2}|0\rangle &-\frac{1}{2}\Delta\varphi 
        & 1 \\\hline
        & \psi^{\Bar{1}}_{-\Delta\varphi-1/2}|0\rangle &\frac{3}{2}\Delta\varphi 
        & 1 \\\hline
        & \psi^{2,\Bar{3}}_{-1/2}|0\rangle
        &\frac{1}{2}\Delta\varphi &\alpha\\ \hline
        & \psi^{\Bar{2},3}_{-1/2}|0\rangle
        &\frac{1}{2}\Delta\varphi & \alpha^2\\ \hline
      \end{array}
&
      \begin{array}{|c||l|c|c||c|} \hline
        \multicolumn{5}{|c|}{\rule[-3mm]{0mm}{8mm} \text{\bf
            Fermionic states on  $T^2 \times T^4/\Z_3$}} \\ \hline\hline
        \text{on }T^2 &  \text{state} &\text{mass}&
        \text{chirality} & \Z_3
\\ \hline\hline
\Delta\varphi=0 & |0\rangle_R & 0 & L &  1\\\hline
& \psi^0_0 \psi^1_0 |0\rangle_R & 0 & R & 1 \\\hline 
& \psi^0_0 \psi^2_0 |0\rangle_R & 0 & R & \alpha \\\hline 
& \psi^0_0 \psi^3_0 |0\rangle_R & 0 & R & \alpha^2 \\\hline 
& \psi^1_0 \psi^2_0 |0\rangle_R & 0 & L & \alpha \\\hline 
& \psi^1_0 \psi^3_0 |0\rangle_R & 0 & L & \alpha^2 \\\hline
& \psi^2_0 \psi^3_0 |0\rangle_R & 0 & L & 1 \\\hline 
& \psi^0_0 \psi^1_0 \psi^2_0 \psi^3_0 |0\rangle_R & 0 & R & 1 \\\hline\hline  
\Delta\varphi \neq 0 & |0\rangle_R & 0 & L & 1 \\\hline
& \psi^0_0 \psi^2_0 |0\rangle_R & 0 & R & \alpha\\\hline 
& \psi^0_0 \psi^3_0 |0\rangle_R & 0 & R & \alpha^2 \\\hline 
& \psi^2_0 \psi^3_0 |0\rangle_R & 0 & L & 1 \\\hline
      \end{array}
 \end{array}
    \end{equation*}

\newpage
\section{Chiral Spectra of Examples 2a and 2b}
\label{app_spectra_2a2b}
\renewcommand{\arraystretch}{1.5}
\begin{table}[h!]
  \begin{center}
    \begin{equation*}
      \begin{array}{|c||c||c||c|c|c|c|c||c|c|c|} \hline
        \multicolumn{11}{|c|}{\rule[-3mm]{0mm}{8mm} \text{\bf
         Chiral fermionic spectrum for example 2a}} \\ \hline\hline
        & \text{mult.} & SU(3) \times SU(2)_L  \times SU(2)_R \times
         SO(8) 
         &Q^1_A & Q^1_B & Q^2_B & Q^1_C & Q^2_C
         & Q_{B-L} & Q^{\prime} & Q^{\prime\prime}
\\ \hline\hline
AB \alpha^0 & 2 & (\Bar{3},2,1,1) & -1 & 1 &0&0&0& 1/3 & 5/3 & 1/4\\
 & 2 & (3,1,2,1) &1 &0&1&0&0& -1/3  & -5/3 & 1/4\\
{} \alpha^1 & 1 & (1,2,1,8) &0&-1&0&0&0& 0 & -1 &-1/4\\
 & 1 & (3,1,2,1) &1&0&-1&0&0& -1/3 & 1/3 &-1/4\\
{} \alpha^2 & 1 & (1,1,2,8) &0&0&-1&0&0& 0 & 1 &-1/4\\
 & 1 & (\Bar{3},2,1,1) &-1&-1&0&0&0& 1/3 & -1/3 &-1/4\\\hline
AC \alpha^0 & 2 & (3,1,1,1) &1&0&0&-1&0& -4/3 & -2/3 & -1/2\\
 & 2 & (\Bar{3},1,1,1) &-1&0&0&0&-1&4/3& 2/3& -1/2\\ 
{} \alpha^1 & 1 & (1,1,1,8) &0&0&0&1&0&1 & 0 & 1/2\\
 & 1 & (\Bar{3},1,1,1) &-1&0&0&0&1& -2/3 & 2/3 & 1/2\\
{} \alpha^2 & 1 & (1,1,1,8) &0&0&0&0&1&-1 & -0 & 1/2\\
 & 1 & (3,1,1,1) &1&0&0&1&0&2/3 & -2/3 & 1/2\\\hline
BB' \alpha^0 & 2 & (1,2,2,1) &0&1&1&0&0&0 & 0& 1/2\\
{} \alpha^1 & 1 & (1,1,1,1) &0&-2&0&0&0&0 & -2 & -1/2\\
{} \alpha^2 & 1 & (1,1,1,1) &0&0&-2&0&0&0 & 2& -1/2\\\hline
CC' \alpha^0 & 2 & (1,1,1,1) &0&0&0&-1&-1&0 & 0 &-1\\
{} \alpha^1 & 1 & (1,1,1,1) &0&0&0&2 &0&2 &0&1\\
{} \alpha^2 & 1 & (1,1,1,1) &0&0&0&0&2 &-2&0&1 \\
\hline
BC' \alpha^0 & 2 & (1,2,1,1) &0&-1&0&-1&0&-1 & -1 &-3/4\\
 & 2 &  (1,1,2,1) &0&0&-1&0&-1&1 & 1&-3/4\\
{} \alpha^1 & 1 & (1,2,1,1) &0&1&0&0&1&-1 &1 & 3/4\\
{} \alpha^2 & 1 & (1,1,2,1) &0&0&1&1&0&1 &-1 & 3/4
\\\hline
      \end{array}
    \end{equation*}
  \end{center}
\caption{Chiral fermionic spectrum for example 2a}
\label{tab_ex2a}
\end{table}

\renewcommand{\arraystretch}{1.5}
\begin{table}[h!]
  \begin{center}
    \begin{equation*}
      \begin{array}{|c||c||c||c|c|c|c||c|c|} \hline
        \multicolumn{9}{|c|}{\rule[-3mm]{0mm}{8mm} \text{\bf
         Chiral fermionic spectrum for example 2b}} \\ \hline\hline
        & \text{mult.} & SU(3) \times SU(2)_L  \times SU(2)_R \times
         SO(8) 
         &Q^1_A & Q^1_B & Q^2_B & Q^1_C
         & Q_{B-L} & Q^{\prime} 
\\ \hline\hline
AB \alpha^0 & 2 & (\Bar{3},2,1,1) & -1 & 1 &0&0& 1/3 & 1 \\
 & 2 & (3,1,2,1) &1 &0&1&0& -1/3  & -1 \\
{} \alpha^1 & 1 & (1,2,1,8) &0&-1&0&0& 0 & -1\\
 & 1 & (3,1,2,1) &1&0&-1&0& -1/3 & 1\\
{} \alpha^2 & 1 & (1,1,2,8) &0&0&-1&0& 0 & 1\\
 & 1 & (\Bar{3},2,1,1) &-1&-1&0&0& 1/3 & -1\\\hline
BB' \alpha^0 & 2 & (1,2,2,1) &0&1&1&0&0 & 0\\
{} \alpha^1 & 1 & (1,1,1,1) &0&-2&0&0&0 & -2\\
{} \alpha^2 & 1 & (1,1,1,1) &0&0&-2&0&0 & 2\\\hline
BC \alpha^0 & 2 & (1,2,1,1) &0&-1&0&1&-1 & 1\\
 & 2 &  (1,1,2,1) &0&0&-1&-1&1 & -1\\
{} \alpha^1 & 1 & (1,1,2,1) &0&0&1&-1&1 & -3\\
{} \alpha^2 & 1 & (1,2,1,1) &0&1&0&1&-1 & 3
\\\hline
      \end{array}
    \end{equation*}
  \end{center}
\caption{Chiral fermionic spectrum for example 2b}
\label{tab_ex2b}
\end{table}

\end{appendix}

\end{document}